\documentstyle[aas2pp4]{article}

\begin{document}
\title{ THE YOUNG AGE OF THE EXTREMELY METAL-DEFICIENT BLUE COMPACT DWARF 
GALAXY SBS~1415+437{\footnote[1]{Based on observations 
obtained with the NASA/ESA {\it{Hubble Space Telescope}} through the Space 
Telescope Science Institute, which is operated by AURA,Inc. under NASA
contract NAS5-26555.}}$^,$\footnote[2]{Ground-based spectroscopic
observations presented herein were obtained with the Multiple Mirror
Telescope, a facility operated jointly by the Smithsonian Institution
and the University of Arizona.}}
\author{Trinh X. Thuan}
\affil{Astronomy Department, University of Virginia, Charlottesville
VA 22903 \\ Electronic mail: txt@virginia.edu}
\author{Yuri I. Izotov{\footnote[3]{Visiting astronomer, 
National Optical Astronomical Observatories.}}}
\affil{Main Astronomical Observatory, Ukrainian National Academy of Sciences,
Goloseevo, Kiev 252650, Ukraine \\ Electronic mail: izotov@mao.kiev.ua}
\and
\author{Craig B. Foltz}
\affil{Multiple Mirror Telescope Observatory, University of Arizona, 
Tucson, AZ 85721 \\ Electronic mail: cfoltz@as.arizona.edu}

\begin{abstract}

We use Multiple Mirror Telescope (MMT) spectrophotometry 
and {\sl Hubble Space Telescope} ({\sl HST}) Faint Object Spectrograph (FOS)
spectra and Wide Field and Planetary Camera 2 (WFPC2) $V$ and $I$ images
to study the properties and evolutionary status of the nearby ($D$ = 11.4 Mpc)
extremely metal-deficient blue compact dwarf (BCD) 
galaxy SBS 1415+437 $\equiv$ CG 389.
The oxygen abundance in the galaxy is 12 + log (O/H) = 7.60 $\pm$ 0.01
or $Z_\odot$/21.

The helium mass fraction in SBS 1415+437 is 
$Y$ = 0.246 $\pm$ 0.004 which agrees with the primordial helium abundance 
determined by Izotov \& Thuan using a much larger sample of BCDs.
The $\alpha$-elements-to-oxygen abundance ratios (Ne/O, S/O, Ar/O) are in
very good agreement with the mean values for other metal-deficient BCDs and 
 are consistent with the scenario that these elements are made in massive stars.
The Fe/O abundance ratio is $\sim$ 2 times smaller than the solar ratio. The
Si/O ratio is close to the solar value, implying that silicon is 
not significantly depleted into dust grains.
The values of the N/O and C/O ratios imply that intermediate-mass 
stars have not had time to evolve in SBS 1415+437 and release their 
nucleosynthesis products and that both N and C in the BCD have been made by 
massive stars only. This sets an upper limit of $\sim$ 100 Myr 
on the age of SBS 1415+437.

 The $V$ and $I$ surface brightness profiles 
of  SBS 1415+437 are well fitted by exponentials implying that
the galaxy is a disk-like system. The velocity distribution derived from the 
H$\alpha$ and [O III] $\lambda$5007 emission lines indicates a solid-body
rotation with a rotational velocity of 80 km s$^{-1}$. The dynamical mass
is $\sim$ 13 times larger than the gas mass, implying that most of the mass 
in SBS 1415+437 is in the form of dark matter.
The $(V-I)$ color of the low-surface-brightness
component of the galaxy is blue ( $\la$ 0.4 mag ) indicative of a very young 
underlying stellar population. The $(V-I)$ - $I$ 
color-magnitude diagrams of the resolved stellar populations
 in different regions of   
 SBS 1415+437 suggest propagating star formation from the NE side
of the galaxy to the SW. 
 
 All regions in SBS 1415+437 possess 
very blue spectral energy distributions (SED). 
By comparing the observed SEDs to theoretical SEDs  
 which include both stellar and gaseous emission, we find that the 
ages of the stellar populations in SBS 1415+437 to range from a few Myr 
 to 100 Myr. Thus chemical abundances, color profiles and 
spectral energy distributions all say that SBS 1415+437 is a truly young galaxy 
which did not start to make stars until $\sim$ 100 Myr ago.

\end{abstract}


\section {INTRODUCTION}

   The evolutionary status of low-metallicity blue compact dwarf (BCD)
   galaxies has been debated
for decades, ever since the pioneering paper by Searle, Sargent \& Bagnuolo 
(1973). 
Are BCDs truly young systems where star formation is occurring for the first 
time or are 
they old galaxies with an old underlying stellar population on which the 
current 
starburst is superposed? An argument in favor of their youth would be their 
 heavy element abundance which ranges between $Z_\odot$/50 and $Z_\odot$/3,
placing them 
among the least chemically evolved galaxies in the universe. On the other hand, 
subsequent surface photometric studies have shown that the majority
 of BCDs are not
young and that they have experienced star formation episodes in the past. 
Loose \& Thuan
(1985) found that $\sim$ 95 \% BCDs in their sample exhibit an underlying red
extended
low surface brightness component, on which are superposed the high
surface brightness star-forming regions. Later CCD surveys of BCDs have
confirmed and strengthened that initial result (Kunth, Maurogordato \& Vigroux
1988; Papaderos et al. 1996; Telles \& Terlevich 1997). 

However, there are
at least two known BCDs with extremely young stellar populations. They are
the two most-metal deficient galaxies known, I Zw 18 ($Z_\odot$/50) and 
SBS 0335--052 ($Z_\odot$/41).
{\sl Hubble Space Telescope} ({\sl HST}) imaging to $V$ $\sim$ 26 of I Zw 18 by 
Hunter \&
Thronson (1995) suggests that the stellar population is dominated by young
stars and that the colors of the underlying diffuse component are consistent 
with those from a sea of unresolved B or early A stars, with no evidence
for stars older than $\sim$ 10$^7$ yr. The BCD SBS 0335--052 was first shown by
Izotov et al. (1990) to possess an extraordinarily low metallicity. {\sl HST}
$V$ and $I$ imaging of this galaxy (Thuan, Izotov \& Lipovetsky 1997)  show
blue $(V-I)$ colors not only in the region of current star formation, but also
in the extended, low-surface-brightness envelope some 4 kpc in diameter. 
Izotov et al. (1997a) and Papaderos et al. (1998) have shown that $\sim$ 1/3
of the emission from 
the underlying component, with color $(V-I)$ = 0.0 - 0.2, comes from
 ionized gas while the remaining 2/3 come from young 
stars not older than $\sim$ 10$^8$ yr. Another piece of evidence in favor of 
the youth 
of SBS 0335--052 is provided by Thuan \& Izotov (1997).
 From {\sl HST} UV
spectrophotometry, they found that SBS 0335--052 is a damped Ly$\alpha$
system with an extremely high neutral hydrogen column density $N$(H I) = 
7 $\times$ 10$^{21}$ cm$^{-2}$, suggesting a large amount of neutral gas
around the galaxy. In fact, a Very Large Array (VLA) map of SBS 0335--052 does 
show the
presence of a large extended H I cloud, 64 kpc by 24 kpc in size, 
and with mass $\sim$ 10$^9$ $M_\odot$, some  
 two orders of magnitude larger than the observed stellar mass 
(Pustilnik et al. 1999). There are two prominent, slightly-resolved H I peaks
 separated by 22 kpc, one of which is associated with SBS 0335--052, and the 
 other with a fainter dwarf galaxy, which is even slightly more metal-deficient 
 ($Z_\odot$/50, Lipovetsky et al. 1999a).
  Thuan \& Izotov (1997) have argued that the 
H I envelope may be truly primordial, unpolluted by heavy elements.

    The above observational evidence suggests that
 there may be truly young dwarf galaxies in the local
 universe which started to make stars for the first time less than about 100 
Myr ago.
 The study of such young galaxies 
 is  important not only for their intrinsic interest,  
but also for understanding galaxy formation at high redshift.
Their proximity allows studies of their structure, metal content and 
stellar populations with a sensitivity, precision and spatial resolution that 
faint, small and 
distant high-redshift galaxies do not allow.
We have therefore started a search for other young galaxies among the
most metal-deficient BCDs known. In that search, we were 
guided by the work of Izotov \& Thuan (1999) who have argued that chemical 
abundances in BCDs can put constraints on their age. Their conclusion, based on 
the behavior of the C/O and N/O ratios as a function of O abundance, 
is that all galaxies with 12 + log (O/H) $\la$ 7.6 ($Z_\odot$/20) began to form 
stars less than $\sim$ 100 Myr ago.

A most attractive candidate is the BCD 
SBS 1415 + 437 $\equiv$ CG 387, which possesses a metallicity of only 
$Z_\odot$/21 and which is near enough ($v$ = 607 km s$^{-1}$) to allow resolution 
into stars. Its general characteristics are shown
in Table 1. At the distance of 11.4 Mpc adopted for SBS 1415+437,
1\arcsec corresponds to 55 pc. We report here new Multiple Mirror 
Telescope (MMT) spectroscopy,
{\sl HST} Faint Object Spectrograph (FOS) UV and optical spectroscopy and 
Wide Field and Planetary Camera 2 (WFPC2) $V$ and $I$ imaging of the BCD. 
We then use these data to discuss the evolutionary status of this nearby dwarf 
galaxy. 
We describe the MMT and {\sl HST} observations in \S2. The heavy element and 
helium abundances are derived in \S3. The morphology and kinematics of the 
ionized
gas are discussed in \S4. In \S5 we discuss the 
 stellar populations in SBS 1415+437. We summarize our findings in \S6.

\section {OBSERVATIONS AND DATA REDUCTION}
\subsection {MMT optical spectroscopy}

Spectrophotometric observations of SBS 1415+437 with a signal-to-noise
ratio S/N = 30 in the continuum were obtained with the Multiple
Mirror Telescope (MMT) on the night of 1997 May 1. 
The observations were made with the Blue Channel of the MMT Spectrograph
using a highly-optimized Loral 3072 $\times$ 1024 CCD detector. A
1\farcs5 $\times$  180\arcsec\ slit was used along with a 500 g/mm grating in
first order and an L--38 second-order blocking filter. This gives a spatial
scale along the slit of 0.3 arcsec\ pixel$^{-1}$, a scale perpendicular
to the slit of 1.9 \AA\ pixel$^{-1}$, a spectral range of
3600--7300 \AA\ and a spectral resolution of $\sim$ 7 \AA\ FWHM. For
these observations, CCD rows were binned by a factor of 2, yielding a
final spatial sampling of 0.6 arcsec\ pixel$^{-1}$.  The observations
cover in a single frame the full spectral range, with all
the lines of interest from [O II] $\lambda$3727 to [O II] $\lambda$7330.
 Furthermore, they have sufficient spectral 
resolution to separate
[O III] $\lambda$4363 from nearby H$\gamma$ and to distinguish between
narrow nebular and broad Wolf-Rayet emission lines. The total exposure time was 
90 minutes and was
broken up into 3 subexposures, 30 minutes each, to allow
for more effective cosmic-ray removal. All exposures were taken at
small airmasses ( $\la$ 1.05 ), so no correction was made for
atmospheric dispersion. The seeing during the observations was 1\arcsec\ 
FWHM.
The slit was oriented in the direction with position
angle 22\arcdeg, along the major axis of the galaxy. 
The spectrophotometric standard star HZ 44 was observed for
flux calibration. Spectra of He--Ne--Ar comparison lamps were obtained
before and after each observation to provide wavelength calibration.

Data reduction of the spectral observations was carried out at the 
NOAO headquarters in Tucson using
the IRAF\footnote[4] {IRAF: the Image Reduction and Analysis Facility is
distributed by the National Optical Astronomy Observatories, which is
operated by the Association of Universities for Research in Astronomy,
In. (AURA) under cooperative argeement with the National Science
Foundation (NSF).} software package. This included bias subtraction,
cosmic-ray removal and flat-field correction using exposures of a
quartz incandescent lamp.
Wavelength calibration was performed for the 2-dimensional spectrum by 
constructing dispersion curves for successive 30 pixels wide regions 
along the slit, using the comparison spectrum.
The amplitude of the scatter of the points 
around each dispersion curve is $\la$ 0.1 \AA.
We then correct the spectrum for distortion and tilt and obtain 
the wavelength-calibrated two-dimensional spectrum by using the 
IRAF routine TRANSFORM.
The accuracy of the wavelength calibration along the slit 
was checked by measuring the wavelength of the bright night sky line 
[O I] $\lambda$5577. We found the dispersion of the measured wavelengths 
to be $\sim$ 0.06 \AA, equivalent to velocity errors of $\sim$ 3 km s$^{-1}$.
 
After night sky background subtraction, and correcting
for atmospheric extinction, each frame was calibrated to absolute fluxes.
To derive the sensitivity curve, we have fitted the observed spectral energy
distribution of the bright hot white dwarf standard star HZ 44 with a 
high-order polynomial. Because the spectrum of HZ 44 has only a 
small number of relatively weak absorption features, its spectral energy 
distribution is known with the very good precision of $\sim$ 1 \% (Oke 1990). We
were thus able to derive a sensitivity curve with a precision of $\sim$ 1 \% 
over the whole 
optical range, except in the region blueward of [O II] $\lambda$3727 
where the sensitivity drops precipitously. 
One-dimensional spectra were extracted by
summing, without weighting, of different numbers of rows along the slit 
depending on the exact region of interest.  
The spectrum of SBS 1415+437 in the region 3600 \AA \ $\leq
\lambda \leq $ 7500 \AA \ for the aperture 1\farcs5 $\times$ 5\arcsec\ is shown in 
Figure 1. The continuum was fitted after removal of the emission lines, and
line intensities were measured by fitting Gaussians to the profiles.

We have adopted an iterative procedure to derive both
the extinction coefficient $C$(H$\beta$) and the absorption equivalent
width for the hydrogen lines simultaneously from the equation (Izotov, Thuan \&
Lipovetsky 1994, 1997, hereafter  respectively ITL94 and ITL97):

%
%
\begin{eqnarray}
\frac{I(\lambda)}{I({\rm H}\beta)}& = &  
\frac{EW_e(\lambda)+EW_a(\lambda)}{EW_e(\lambda)} 
\frac{EW_e({\rm H}\beta)}{EW_e({\rm H}\beta)+EW_a({\rm H}\beta)}\times \nonumber \\
& & \frac{F(\lambda)}{F({\rm H}\beta)}10^{[C({\rm H}\beta)f(\lambda)]}, \label{eq:CHb} 
\end{eqnarray}
where $I$($\lambda$) is the intrinsic line flux and $F$($\lambda$) is the
observed line flux corrected for atmospheric extinction.
$EW$$_e$($\lambda$) and $EW$$_a$($\lambda$) are the equivalent widths of
the observed emission line and the underlying absorption line,
respectively, and $f$($\lambda$) is the reddening function, normalized at
H$\beta$, which we take from Whitford (1958).

We use the theoretical ratios from Brocklehurst (1971) at the electron
temperature estimated from the observed [O III]($\lambda$4959 +
$\lambda$5007)/$\lambda$4363 ratio for the intrinsic hydrogen line
intensity ratios. For lines other than  hydrogen $EW$$_a$($\lambda$) = 0, so
Eq. (\ref{eq:CHb}) reduces to
 
\begin{equation}
\frac{I(\lambda)}{I({\rm H}\beta)} = \frac{F(\lambda)}{F({\rm H}\beta)} 
10^{[C({\rm H}\beta)f(\lambda)]}. \label{eq:CHb1}
\end{equation}

The observed and corrected line intensities, extinction coefficient, 
and equivalent widths of
the stellar hydrogen absorption lines are given in Table 2 along with
the uncorrected H$\beta$ flux and H$\beta$ equivalent width for the brightest
part of SBS 1415+437. We give the line intensities for two different apertures, 
a large one (1\farcs5 $\times$ 5\arcsec) and a small one 
(1\farcs5 $\times$ 0\farcs6). 
While the spectrum associated with the larger aperture has a higher 
signal-to-noise ratio, 
the one associated with the smaller aperture is useful for comparing with the 
{\sl Hubble Space Telescope} ({\sl HST}) Faint Object
Spectrograph (FOS) spectra which were obtained with a 0\farcs86 circular
aperture.

\subsection{{\sl HST} observations}

\subsubsection{Imaging}

  We obtained images of SBS 1415+437 on 1995 April 17 during cycle 4, after the 
  refurbishment mission,
  with the {\sl HST} Wide Field and Planetary Camera 2
(WFPC2) in filters F569W and 
F791W, which we will refer to as $V$ and $I$ throughout the paper.
Two exposures of equal duration were obtained in each filter to permit 
identification and 
removal of cosmic rays. 
The total exposure time was 1800s in $V$ and 4400s in $I$. The scale of the 
WFPC2
is 0\farcs102 per pixel. 

   Preliminary processing of the raw images including corrections for
flat-fielding was done at the Space Telescope Science Institute through the 
standard 
pipeline. Subsequent reductions were carried out at the University of Virginia
using IRAF
and STSDAS{\footnote[5]{STSDAS: the Space Telescope Science Data Analysis System.}}.
 Cosmic rays were removed and the images 
in each filter were combined. We found that all exposures in a given filter 
coregistered to better than $\sim$ 0.2 pixels.
The transformation of instrumental magnitudes to the Johnson-Cousins $UBVRI$ 
photometric system as defined by Landolt (1992) was performed according to the 
prescriptions of Holtzman et al. (1995b).


 The resulting measured
brightness of the sky background is 23.0 mag arcsec$^{-2}$ in $V$ and
22.2 mag arcsec$^{-2}$ in $I$.

\subsubsection{Spectrophotometry}

    We also obtained ultraviolet and optical spectra of SBS 1415+437 with the 
Faint Object Spectrograph (FOS) during cycle 4 on 1995 January 24.
The UV spectrum was obtained with grating G190H and an integration 
time of 1600s and covers the wavelength range $\lambda$1572 -- 2312 \AA. 
The optical spectra were obtained with grating G400H and an exposure time of 
960s in 
the blue range ($\lambda$3235 -- 4781 \AA) and with grating G570H and an 
exposure time of 1260s in the red range ($\lambda$4569 -- 6818 \AA). 
The spectra were taken consecutively in time with the 0\farcs86 
circular aperture 
(aperture B-3) giving a spectral resolution of $\sim$ 3 \AA\ for the UV 
spectrum,
$\sim$ 6 \AA\ for the blue spectrum and $\sim$ 9 \AA\ for the red one.

  The spectra were processed through the standard {\sl HST} 
pipeline reductions, 
with corrections for flat-fielding, wavelength determination, instrumental 
background subtraction and photometric calibration. The instrumental and 
sky backgrounds were negligible for our observations. The FOS spectra are
shown in Fig. 2 where all emission lines are marked. Note that in the UV
spectrum, the Si III] $\lambda$1892 emission line is not present. Instead,
there is an absorption feature at the location of the line. 
By examining the count rates vs diodes in the uncalibrated spectrum, we 
have determined the absorption feature to be caused by several bad diodes.  
Thus for our analysis,  we shall assume that the Si III] $\lambda$1892 
emission line is emitted in the H II region with a flux 0.67 times that 
of Si III] $\lambda$1883, as
expected in the low density limit from the statistical weight ratio of the
Si$^{2+}$ $^3$P$_2$ and $^3$P$_1$ levels.

   The observed emission line fluxes in the FOS spectra are dereddenned
according to  
Eq. (\ref{eq:CHb}) and (\ref{eq:CHb1}). In the optical range we again use the reddening function
from Whitford (1958), and in the UV range the Small Magellanic Cloud reddening
curve as parameterized by Pr\'evot et al. (1984).

The observed and corrected line fluxes of the {\sl HST} spectra are shown
in Table 2 together with those from the ground-based optical observations.
The errors given include statistical uncertainties plus a 1 \% error 
in the flux calibration. They have been propagated to derive  
element abundances and their errors.
The extinction coefficients $C(H\beta)$ derived from the MMT small aperture
spectrum and the {\sl HST} spectrum are very similar. They are 
very small, being 0.025 and 0.020 dex, respectively.

\section{HEAVY ELEMENTS AND HELIUM ABUNDANCES}

Since SBS 1415+437 has been observed over a wide spectral range with a 
high signal-to-noise ratio, 
reliable heavy element abundances of the ionized gas
can be derived. The heavy element 
abundance ratios allow to place constraints 
on stellar nucleosynthesis processes in a very metal-deficient 
environment and on the chemical evolution of the BCD.
Of special interest is
the determination of the helium abundance which should approximate well  
the primordial value as SBS 1415+437 is one of the most metal-deficient BCDs
 known. 

\subsection{Abundance determination}

To derive element abundances from the optical line intensities,
 we follow the procedures detailed in ITL94 and ITL97. It is known that the 
electron
temperature, $T_e$, is different in high- and low-ionization zones of H II
regions (Stasi\'nska 1990). We have chosen to determine $T_e$(O III) from the
[O III] $\lambda$4363/($\lambda$4959+$\lambda$5007) ratio 
using the five-level atom model (Aller 1984)  and $N_e$(S II) from
the [S II] $\lambda$6717/$\lambda$6731 ratio. 
We adopt $T_e$(O III) for the 
derivation of He$^+$, He$^{2+}$, O$^{2+}$, Ne$^{2+}$ and Ar$^{3+}$ ionic 
abundances. To derive the electron temperature for the O$^+$ ion, we have used 
the relation between
$T_e$(O II) and $T_e$(O III) (ITL94), based on the
photoionization models of Stasi\'nska (1990). $T_e$(O II) has been used to
derive the O$^+$, N$^+$ and Fe$^+$ ionic abundances. For Ar$^{2+}$
and S$^{2+}$ we have used an electron temperature intermediate between
$T_e$(O III) and $T_e$(O II) following the prescriptions of Garnett (1992).

Total element abundances have been derived after correction for unseen stages
of ionization as described by ITL94.
The spectrum of SBS 1415+437 shows a strong nebular He II $\lambda$4686
emission line, implying the presence of a non-negligible amount of O$^{3+}$. 
Its abundance is derived from the relation:
\begin{equation}
{\rm O}^{3+}=\frac{{\rm He}^{2+}}{{\rm He}^+}({\rm O}^++{\rm O}^{2+}). 
\label{eq:O3+}
\end{equation}
Then, the total oxygen abundance is equal to
\begin{equation}
{\rm O}={\rm O}^++{\rm O}^{2+}+{\rm O}^{3+}.       \label{eq:O}
\end{equation}
The electron temperatures, number densities, ionic and total element
abundances along with ionization correction factors are shown in Table 3.

To derive carbon and silicon abundances we adopt the electron temperature
obtained from the MMT small aperture spectrum as the latter has a 
much better signal-to-noise ratio than the FOS spectrum. This is 
legitimate as 
the area of the small square aperture of the MMT spectrum is only 1.55 times 
 that of the FOS aperture, similar to
the ratio 1.47 of the H$\beta$ fluxes in the two apertures.

We derive the C$^{2+}$ abundance from the relation (Aller 1984, p.125-126):
\begin{equation}
\frac{{\rm C}^{2+}}{{\rm O}^{2+}}=0.093\exp\left(\frac{4.656}{t}\right)
\frac{I({\rm C III]}\lambda1906+\lambda1909)}
{I({\rm [O III]}\lambda4959+\lambda5007)},
\label{eq:C2+}
\end{equation}
where $t$ = $T_e$/10$^4$. Following Garnett et al. (1995a) we adopt
for the temperature in Eq. (\ref{eq:C2+}) the value $T_e$(O III) =
16600 $\pm$ 270 K derived from the spectrum in the small MMT aperture.
As the O III] $\lambda$1666 emission line is not seen, we cannot
derive an extinction-independent C$^{2+}$/O$^{2+}$ ratio. However, 
the extinction derived from the FOS spectrum is very small
( $C$(H$\beta$) = 0.02 dex ). Assuming $C$(H$\beta$) = 0 we 
derive C$^{2+}$/O$^{2+}$, about 5 \% lower than the 
value in Table 3. Adopting 
$T_e$(O III) = 17100 $\pm$ 2700 K as derived from
the noisier FOS spectrum we obtain a very similar C$^{2+}$/O$^{2+}$ ratio. 
We calculate the correction factor ICF(C) for unseen ionization stages of 
carbon 
following Garnett et al. (1995a). In the case of SBS 1415+437 the
correction is small, ICF(C) being equal to 1.05.

The silicon abundance is derived following Garnett et al. (1995b) from the
relation
\begin{equation}
\frac{\rm Si}{\rm C}={\rm ICF}({\rm Si})\frac{{\rm Si}^{2+}}{{\rm C}^{2+}},
\label{eq:Si}
\end{equation}
where
\begin{equation}
\frac{{\rm Si}^{2+}}{{\rm C}^{2+}}=0.188t^{0.2}\exp\left(\frac{0.08}{t}\right)
\frac{I({\rm Si III]}\lambda1883+\lambda1892)}
{I({\rm C III]}\lambda1906+\lambda1909)}.
\label{eq:Si2+}
\end{equation}
For the determination of the silicon abundance 
we again adopt the temperature $T_e$(O III) = 16600  $\pm$  270 K. 
The correction factor ICF(Si) = 1.33 is derived
following Garnett et al. (1995b). As discussed before, we have assumed
$I$(Si III] $\lambda$1883 + $\lambda$1892) = 
1.67  $\times$  $I$(Si III] $\lambda$1883) since the Si III] $\lambda$1892
emission line is not seen, although the
signal-to-noise ratio of the spectrum and
the intensity of the Si III] $\lambda$1883 emission line 
imply that it should be present.

\subsection{Heavy element abundances and age constraints}

The heavy element abundances in SBS 1415+437 corresponding to
the two MMT apertures 
and those corresponding to the {\sl HST}  FOS aperture are shown in Table 3. 
They are in very good 
general agreement with the values derived by Izotov \& Thuan (1998b, 1999) 
from 4-meter Kitt Peak telescope observations. In particular, the oxygen
abundance derived by Izotov \& Thuan (1999) is 12 + log (O/H) = 7.59  $\pm$  0.01.
The ratios of $\alpha$-process elements Ne, S, Ar abundances to
O abundance are in very good
agreement with the mean values derived for low-metallicity blue compact 
galaxies
by Thuan, Izotov \& Lipovetsky (1995, hereafter TIL95) and Izotov \& Thuan 
(1999).

We find that the N/O ratios derived from both MMT spectra (log (N/O) =
--1.54  $\pm$  0.03 and --1.55  $\pm$  0.03 ) are in agreement with the mean value
log (N/O) = --1.53  $\pm$  0.08 derived by TIL95 for their sample of very 
metal-deficient BCDs, but are slightly higher than the value --1.58  $\pm$  0.02 
derived by Izotov \& Thuan (1998b) for this galaxy from 4-meter Kitt Peak telescope 
observations.
The small difference may be due to a larger contamination  in 
the lower spectral resolution MMT spectrum of the [N II] $\lambda$6584
emission line by the strong nearby H$\alpha$ $\lambda$6563 emission line.  
In spite of these small subtle differences, all determinations of the N/O ratio
in SBS 1415+437 are in agreement within the errors.
Hence, SBS 1415+437 falls within the narrow range of N/O ratios found in 
 other low-metallicity BCDs. 
 For BCDs with 12 + log (O/H) $\la$ 7.6, TIL95 and Izotov \& Thuan (1999)
found that the dispersion 
about the mean of the N/O ratio is remarkably small, being only 0.02 
dex. The constancy of N/O with O abundance implies that N and O share 
the same origin. Moreover, the very small dispersion is strong evidence 
against time-delayed production of primary N and supports the view 
 that primary N in galaxies with $Z$ $\la$ $Z_\odot$/20 is  
 produced by massive ($M$ $>$ 9 $M_\odot$) stars only. Intermediate-mass
(3 $M_\odot$ $\la$ $M$ $\la$  9 $M_\odot$) stars 
in those galaxies have not had time 
enough to evolve to contribute primary N. Hence, the observed N/O ratio
 in SBS 1415+437 implies that it is a young galaxy with an age not 
exceeding $\sim$ 100 Myr (the lifetime of a 9 $M_\odot$ star being $\sim$ 40 
Myr).

Another important age discriminator is the Fe/O ratio. While oxygen is
produced only by massive stars, production of iron occurs not only during
explosive nucleosynthesis in massive stars, but also in SNe I about 1 Gyr after
the beginning of star formation in the BCD. In the latter case a nearly solar
Fe/O ratio is expected. However, TIL95 and Izotov \& Thuan (1999)
 have shown that at low metallicities (12 + log (O/H) $\la$ 8.2),
O is overproduced with respect to Fe as compared to the solar
neighbohrhood. The mean value [O/Fe] (= log (O/Fe) -- log 
(O/Fe)$_\odot$)
for low-metallicity BCDs is 0.32  $\pm$  0.11 (Izotov \& Thuan 1999)
 and compares well with values
observed in Milky Way halo stars (TIL95). In SBS 1415+437, 
we find [O/Fe] = 0.27  $\pm$  0.03 and
0.30  $\pm$  0.07 respectively 
for the large and small apertures. These values
are consistent within the errors with the mean value for low-metallicity BCDs, 
although
the value for the small aperture is more uncertain due to the weakness of 
[Fe III] $\lambda$4658 emission line and its possible contamination by
the Wolf-Rayet C IV $\lambda$4658 line. The second possibility 
is less likely however, as no other broad
Wolf-Rayet lines are seen in the small aperture spectrum. The high [O/Fe] value
seen in SBS 1415+437 as compared to the solar neighborhood  
 also suggests a small age for this galaxy, although it can be explained, in 
principle, by depletion of iron into dust grains.

Further chemical constraints on the age of SBS 1415+437 come from {\sl HST}
FOS spectroscopy. We show the C/O abundance ratio vs. oxygen
abundance in Figure 3a and the Si/O abundance ratio vs. oxygen abundance
in Figure 3b. The filled circle represents SBS 1415+437 while open circles 
represent data from Izotov \& Thuan (1999).
The silicon-to-oxygen abundance ratio 
in SBS 1415+437 (log (Si/O) = --1.46  $\pm$  0.27) agrees well with the solar 
value (log (Si/O)$_\odot$ = --1.38).  It is independent of O abundance  
as expected since both Si and O have
the same origin: they are both produced by high-mass stars. Although errors in
the determinations of the Si abundances are still large, there is no
evidence for significant depletion of
silicon into dust grains as suggested by Garnett et al. (1995b).
This would argue for no depletion of iron in SBS 1415+437 as well.

While silicon is produced only by high-mass stars, carbon is a primary element
produced by both intermediate and high-mass stars. 
Therefore, the C/O ratio is a sensitive indicator of
 the age of a galaxy (Izotov \& Thuan 1999). 
Figure 3a shows that the value of log(C/O) = --0.78  $\pm$  0.10 
in SBS 1415+437 is in good agreement with those
derived for other low-metallicity BCDs with 12 + log (O/H) $\la$ 7.6
(Izotov \& Thuan 1999). 
 The observed log (C/O)
in SBS 1415+437 is consistent with the value of $\sim$ --0.83 predicted by
the theory of massive star nucleosynthesis (Weaver \& Woosley 1993; Woosley
\& Weaver 1995) and with log (C/O) observed in halo stars (Tomkin et al. 1992).
This suggests that the carbon seen in the BCD has been made by 
high-mass stars only, and that intermediate-mass stars have not had 
enough time to 
evolve and add their own C contribution. As in the case of N/O, the value 
of the C/O ratio allows us to say that the first star formation in SBS 1415+437
has not occurred more than $\sim$ 100 Myr ago.
 
 In summary, the analysis of the relative heavy element abundance ratios
  in SBS 1415+437
has led us to an important conclusion: all observed species have been made 
by massive stars. The absence of evidence of any element production
by intermediate-mass stars allows to put an age upper limit 
of $\sim$ 100 Myr for the BCD. 

\subsection{Spatial distribution of heavy element abundances}

The spatial distribution of element abundances provides important 
constraints on  the  
 enrichment and mixing processes in the interstellar medium of galaxies. 
Kobulnicky \& Skillman (1996) discussed this problem 
for the metal-rich dwarf galaxy NGC 4214,
and show that  inside H II regions the material is well mixed and
is characterized by nearly constant element abundances. In low-metallicity
environments, such studies have been carried out by Martin (1996) for I Zw 18 
and by Izotov et al. (1997a) for SBS 0335--052. The results were very similar. 
Martin (1996)
found the oxygen abundance to be nearly 
constant inside the central 530 pc of I Zw 18 and to be within 20 \% of 
the abundance in the 
NW H II region (12 + log (O/H) = 7.1 -- 7.3). She argued that the detection of 
a superbubble in I Zw 18 implies 
a timescale of $\sim$ 15 -- 27 Myr and a spatial scale of $\sim$ 900 pc 
for dispersing the recently synthesized elements. 
Izotov et al. (1997a) found the 
constancy of the heavy element abundances to hold within the errors at 
even larger scales in SBS 0335--052, over the central 2 kpc 
around the youngest cluster.

 Large abundance variations have been seen in only two
galaxies thus far. 
Walsh \& Roy (1987, 1989) found the N/O ratio in different knots in 
the starburst galaxy NGC 5253 to vary by factors of up to 3 and that 
these variations are associated with the presence of Wolf-Rayet stars of WN 
type. A more extreme case is that of the BCD Mkn 996 
(Thuan, Izotov \& Lipovetsky 1996) where the
N/O ratio in the extremely compact central part of the galaxy is a factor of 
$\sim$ 25 greater
than the mean value found in BCDs, and decreases steeply outward.
Local nitrogen enrichment in Mkn 996 is associated with WN stars as well. 
Both these galaxies are however relatively more metal-rich
($Z$ $\ga$ $Z_\odot$/10) and the nuclear H II region in Mkn 996 is some five 
orders of magnitude denser 
 (its $N_e$ is $\sim$ 10$^7$cm$^{-3}$) than typical H II regions in BCDs.
 
SBS 1415+437 is on the other hand a more typical low-metallicity BCD. 
Figure 4 shows the spatial variations of the electron temperature, oxygen
abundance, and abundance ratios of various heavy elements relative to oxygen
along the major axis of the galaxy. Data for two distinct H II regions are 
presented,
that for the brightest H II region at the SW edge of the SBS 1415+437 and that 
for the H II region at 16\arcsec\ ( $\sim$ 880 pc ) 
in the NE direction from it.  Although [O III] $\lambda$5007 and H$\alpha$ 
emission 
are observed everywhere between these two H II regions, the emission lines 
are too weak for abundance measurements. The abundances are derived from 
one-dimensional spectra extracted from the MMT two-dimensional spectrum.
The auroral [O III] $\lambda$4363 emission line is detected only in the 
brightest region, $\sim$ 5\arcsec\  along the slit, allowing a direct 
determination of
$T_e$(O III). In the spectra where the [O III] $\lambda$4363 emission
line is not seen, we use the approximation
\begin{equation}
t_e({\rm O III}) = 2.63-1.37\times\log
\left[\frac{I(\lambda3727+
\lambda4959+\lambda5007)}{I({\rm H}\beta)}\right],
\label{eq:Te}
\end{equation}
where $t_e$ = $T_e$/10$^4$.
The above equation was derived using the sample of
 low-metallicity BCDs 
of ITL94, ITL97 and TIL95 with 12 + log (O/H) 
$\la$ 7.9 and  
precise electron temperature determinations
(the mean error in $T_e$ is $\sim$ 1000 K).

Inspection of Figure 4 shows that the electron temperature and element
abundances within the brightest H II region are
remarkably constant with radius, out to $\sim$ 350 pc.
The fainter H II region appears to show a slight decrease in 
oxygen abundance at the furthest point, when 12 + log (O/H) drops below 7.4, 
although the difference is not significant. We conclude that in SBS 1415+437 
just as in SBS 0335--052, the elemental abundances
do not vary within the errors, although the 
spatial scales probed are smaller. 
In subsection 3.2, we have concluded that the H II regions in 
SBS 1415+437 are most likely enriched by the products of nucleosynthesis in
massive stars. Since the lifetime of massive stars is in the range 
10$^6$ -- 10$^7$ yr, the time scale for the enrichment of the ionized gas with 
heavy elements is also of the same order. The constancy of the heavy element 
abundances is consistent with a scenario of self-enrichment of the 
gas followed by rapid mixing. A scenario of self-enrichment of  
giant H II regions has been proposed by Kunth \& Sargent (1986), who suggested 
that new heavy element ejecta from supernovae of type II and stellar winds  
initially mix exclusively with the ionized gas in the H II zone, while further 
mixing with the cold gas happens only later during the long interburst phase. 
Roy \& Kunth (1995) have analyzed different
mechanisms of interstellar medium mixing and found that the ionized gas is well
mixed due to Rayleigh-Taylor and Kelvin-Helmholz instabilities 
on time scales of $\sim$ 1.5 $\times$ 10$^6$ yr within regions of $\sim$ 100 pc 
size. The enriched gas in the H II regions can be easily transported by 
expanding shells of supernovae and stellar winds in a short time scale. 
Such a picture is also supported by the observation of  
ionized gas velocities of order 10$^3$ km s$^{-1}$ in several
dwarf galaxies (Roy et al. 1992, Skillman \& Kennicutt 1993 and Izotov et al. 
1996). The presence of several superbubbles in the WFPC2 images of SBS 1415+437 
suggest that these fast motions are likely to be also present in the BCD.

\subsection{Helium Abundance}

The extremely low oxygen abundance of SBS 1415 + 437 ( $Z_\odot$/21 ) 
as well as 
the high surface brightness of its star-forming region implies that the galaxy 
is one of the best objects for primordial helium abundance determination. The 
high S/N
ratio spectrum obtained for SBS 1415+437 (Figure 1) allows us to measure
helium line intensities with good accuracy. The He I line intensities
corrected for interstellar extinction are shown in Table 2 for both the large
and small apertures. 
For helium abundance determination we use the He I $\lambda$4471, 
$\lambda$5876 and $\lambda$6678 emission lines. However, the He I line 
intensities deviate from their pure recombination values and must be corrected 
for several mechanisms (ITL97, Izotov \& Thuan 1998ab).
 One of the main mechanism is collisional excitation from the 
metastable 2$^3$S level. This mechanism depends on the electron temperature and 
electron number density. The line most sensitive to 
collisional enhancement in the optical range is He I $\lambda$7065.
Another mechanism is self-absorption in some optically thick emission lines,
such as He I $\lambda$3889. The emission lines most sensitive to this 
fluorescence mechanism are He I $\lambda$3889 and $\lambda$7065. 
In contrast to collisional enhancement, which increases the 
intensities of all He I lines, the fluorescence mechanism works in such a way 
as to decrease the intensity of the He I $\lambda$3889 line as its
optical depth increases, while increasing the intensities of other lines of
interest (He I $\lambda$4471, $\lambda$5876, $\lambda$6678 and $\lambda$7065).
And, finally, underlying He I stellar absorption lines produced by O and B 
stars which decrease the He I emission line intensities, has to be taken into 
account. 
To correct the He I emission line intensities for collisional and 
fluorescent enhancement we follow the approach described by ITL97 and Izotov \& 
Thuan (1998b). 
We assume that the electron temperature in the He$^+$ zone is 
equal to that in the O$^{2+}$ zone. However, we do not set its 
electron number density equal to that derived from the [S II] 
$\lambda$6717/$\lambda$6731 line ratio  
for two reasons:
(1) the S$^+$ and He$^+$ zones do not coincide; 
(2) although an electron number density $N_e$(S II) = 80 cm$^{-3}$ was derived
in the large aperture, a much smaller value is obtained in the small aperture. 
We adopt the value of 10 cm$^{-3}$. Both $N_e$(S II) values are 
small and subject
to large uncertainties because the sulfur emission
line intensity ratio [S II] $\lambda$6717/$\lambda$6731 is not sensitive
to $N_e$ at these low number densities. Therefore we use the intensities of the 
5 He I  
$\lambda$3889, $\lambda$4471, $\lambda$5876, $\lambda$6678 and $\lambda$7065
emission lines to solve self-consistently for 1) the electron number
density in the He$^+$ zone $N_e$(He$^+$) and 2) the optical depth 
$\tau$($\lambda$3889) in the He I $\lambda$3889 line, to reproduce the 
theoretical He I line intensity recombination ratios.

The best solution for both apertures gives $N_e$(He$^+$) = 10 cm$^{-3}$ and 
$\tau$($\lambda$3889) = 0.1. 
The helium abundance for individual lines along with their 
correction factors for collisional and fluorescent enhancement are given in
Table 4. The helium mass fraction in SBS 1415+437, $Y$ = 0.246 $\pm$ 0.004
for the large aperture and $Y$ = 0.245 $\pm$ 0.005 for the small aperture, 
compares well with the value 0.244 $\pm$ 0.002 derived by Izotov \& Thuan (1998b)
for this galaxy. It is in
good agreement with that derived for other low-metallicity 
BCDs and is close to the primordial helium mass fraction $Y_p$ = 
0.244 $\pm$ 0.002 
derived by Izotov \& Thuan (1998b). Note that in general, the corrections to 
the He I line intensities are small because the H II region in SBS 1415+437 is 
not so hot and dense as that in the more metal-deficient BCD SBS 
0335--052 (Izotov et al. 1997a). The H$\beta$ equivalent widths of 159 \AA\ for 
the large aperture and 140 \AA\ for the small aperture, are large compared to 
the value of 56 \AA\ measured in the NW component of I Zw 18 (Izotov et al. 
1997b). Therefore, the equivalent widths of the He I emission lines are
larger as well and underlying stellar absorption does not play as important a 
role as in the case of I Zw 18.

\section{MORPHOLOGY AND KINEMATICS OF THE IONIZED GAS}

    We display in Figure 5 the $I$ image of SBS 1415+437 with the 
contrast level adjusted so as to show the
low-surface-brightness underlying extended component. The galaxy has an 
elongated shape with a bright H II region at its SW tip 
and belongs to the class of ``cometary galaxies'' in the BCD classification 
scheme of Loose \& Thuan (1985).  Many
point sources are seen which are identified as luminous stars. To the SW
of brightest H II region, two stellar clusters with resolved stars are
present. Although we have not observed these two clusters spectroscopically, 
they are likely at the same distance as the main body of the
galaxy, since they are connected to it by diffuse low-surface-brightness light. 
The rest of the light comes from an unresolved lower surface brightness 
extended stellar component. The magnified $I$ image of the bright H II 
region (Figure 6)  shows resolved luminous stars which are nearly aligned,
implying propagating star formation. The structure of the  
star-forming region in SBS 1415+437 is different from that observed in
SBS 0335--052 where several luminous compact super-star clusters are seen 
(Thuan, Izotov \& Lipovetsky 1997). Instead, star-formation in
SBS 1415+437 is more quiescent and is more similar to what is observed
in I Zw 18 (Hunter \& Thronson 1995). It appears that the mode of star 
formation varies in different BCDs. The BCD's total mass may play a role: SBS 
0335--052 where star formation occurs in super-star clusters has a H I mass 
$\sim$ 6 times greater than that in SBS 1415+437 (Thuan et al. 1999a).  

In Figure 7 we show the $(V-I)$ color map of SBS 1415+437, where dark denotes 
blue colors and light denotes red colors. The 
very blue H II regions in the SW tip can be seen, together with  
extended emission with a moderately blue color and red supergiant stars 
(white points) along the main body of the galaxy. The southernmost of the two 
H II regions seen to the SW of the brightest H II region is very blue, while in 
the other H II region red supergiants are already present. The structure of the 
brightest H II region in the $(V-I)$ image is very filamentary, as seen better  
in the magnified picture (Figure 8). This filamentary structure is caused in 
part by the presence of supershells of ionized gas delineating supernova 
cavities, and in part by the presence of dust 
patches, as evidenced by the extinction coefficient $C$(H$\beta$) = 0.12 dex  
derived from the spectrum in the large aperture. Figure 9 shows that there is a 
systematic reddening of the $(V-I)$ color for the stars in the chain seen in 
Figure 6, away from the star at the southernmost tip, implying propagating star 
formation from the NE to the SW.  

A long-exposure long-slit MMT spectrum shows that [O III] 
$\lambda$5007 and H$\alpha$ emission are present along most of the body of 
SBS 1415+437 which has an angular diameter of 46\arcsec\ at the surface 
brightness isophote of 25 mag arcsec$^{-2}$. Only the outermost 13\arcsec\ of 
the galaxy at the NE side do not show emission lines. We measure the  
wavelength shifts of these two lines to derive radial velocities and a rotation 
curve.
The velocity distribution along the major axis of SBS 1415+437 is shown in
Figure 10. As the velocities derived from each of the lines do not show 
significant systematic relative shifts, we have fitted  
a linear relation to the data, between the radial velocity and the location 
along the slit, using a maximum likelihood method.
The origin in distance is taken to be at the maximum of the brightness 
distribution. The resulting
linear regression is shown by the solid line. The velocity 
profile clearly indicates solid-body rotation in SBS 1415+437, 
with a rotational velocity of $\sim$ 50 km s$^{-1}$ over the part of the galaxy 
with emission lines, i.e. a region with a spatial
angular extent of $\sim$ 30\arcsec. Extrapolation of the rotation curve to the 
whole diameter of 46\arcsec\  gives a rotational velocity
$v_{rot}$ = 80 km s$^{-1}$. This yields a dynamical mass
 $M_T$ = $G^{-1}$ $r$ $v_{rot}^2$ $\sim$ 1.9 $\times$ 10$^9$ $M_\odot$.   
Since the H I mass is 1.48 $\times$ 10$^8$ $M_\odot$ (Thuan et al. 1999a), 
this gives 
a $M$(H I)/$M_{T}$ ratio of 0.08, in the range of 
 the values obtained 
in other BCDs such as I Zw 36 (0.14, Viallefond \& Thuan 1983) or I Zw 18 (
0.10, van Zee et al. 1998). As the mass of the visible stellar population 
does not exceed 
$\sim$ 10$^6$ $M_\odot$ (see section below),   
a large part of the mass of SBS 1415+437 is dark.

\section{STELLAR POPULATIONS}

\subsection{Surface brightness and color distributions}

   Surface photometry was done by fitting ellipses to isophotal contours
using the task ELLIPSE in STSDAS.
We show in Figure 11a the $V$ and $I$ surface brightness distributions
as a function of $r$, the equivalent radius of the best fitting elliptical 
isophote defined as $r$=$\sqrt{ab}$, where $a$ and $b$ are respectively
the semi-major and semi-minor axes of the ellipse. 
Both $V$ and $I$ surface brightness
profiles (SBP) can be fitted for $ r\geq 4''$ by an exponential law of the form 
$I = I_0\exp(-\alpha)$, characteristic of a disk structure.   
The best fits to the outer parts are shown by straight lines in 
Figure 11a and are given by:
\begin{equation}
\mu_V = (21.00\pm0.04) + (0.205\pm0.005)r,    \label{eq:muV}
\end{equation}
\begin{equation}
\mu_I = (20.69\pm0.03) + (0.197\pm0.004)r.    \label{eq:muI}
\end{equation}
This corresponds to scale-lengths $\alpha^{-1}_V$ = $\alpha^{-1}_I$ =
0.27 kpc, in the lower range of the values found for BCDs 
(Papaderos et al. 1996, Lipovetsky et al. 1999b).
 The central surface brightnesses are $\mu_{V_0}$ = 21.00 mag\ 
arcsec$^{-2}$ and $\mu_{I_0}$ = 20.69 mag\ arcsec$^{-2}$ as Galactic 
extinction is negligible (Burstein \& Heiles 1982).  
The rise above the exponential law for $r\leq3''$ is caused by the bright  
star-forming region.

Figure 11b shows the ($V-I$) color profile of SBS 1415+437. 
The very blue color at the center ($(V-I)$ $\la$ 0.0 for $r$ $\la$ 2$\arcsec$) 
is due to the bright star-forming region.
The galaxy becomes gradually redder with increasing $r$ reaching a constant 
($V-I$) $\sim$ 0.4 at $r\geq$ 5\arcsec. This constant color is that of the 
underlying low-surface-brightness component.
The $V_{25}$ isophotal radius at the
surface brightness level of 25 mag arcsec$^{-2}$ is 16\arcsec, 
corresponding to a linear radius of 0.88 kpc. 
Total magnitudes inside the 25 mag arcsec$^{-2}$ isophotal level are
$V$ = 15.55 $\pm$ 0.01 mag and $I$ = 15.47 $\pm$ 0.01 mag.
Adopting $B$ = 15.60 mag (Lipovetsky et al. 1999b), we derive 
$L_B$ = 1.2 $\times$ 10$^8$ $L_{B\odot}$, giving 
($M_{\rm H I}/L_B$)/($M/L_B$)$_\odot$ = 1.3 and 
($M_T/L_B$)/($M/L_B$)$_\odot$ = 16.
 These ratios are relatively 
high compared the respective mean values of 0.6 and 5 obtained by 
Thuan (1985) for a large sample of BCDs. This indicates that a relatively small 
fraction of the gas in SBS 1415+437 has been converted into stars and supports 
the conclusion that the BCD is young.

The blue ($V-I$) color ($\sim$ 0.4 mag) of the extended underlying
low-surface-brightness stellar component is remarkable. 
This color is as blue as the ($V-I$) color corrected for gaseous emission 
of the underlying component 
of SBS 0335--052 which Thuan et al. (1997) and Izotov et al. (1997a) 
have argued to be a young galaxy, with stars not older than $\sim$ 100 Myr.
Comparison with evolutionary synthesis models
(Leitherer \& Heckman 1995; Tantalo et al. 1996) also gives an upper
limit for the age of the stellar populations in SBS 1415+437 of $\sim$ 100 Myr.
These models predict a ($M/L_B$)/($M/L_B$)$_\odot$ ratio of $\sim$ 0.01 for an age of 
100 Myr, giving a stellar mass of 1.2 $\times$ 10$^6$ $M_\odot$.   
Direct age determinations for the stellar populations in SBS 1415+437 
can be obtained from color-magnitude diagrams which we discuss next.


\subsection{Color-magnitude diagrams}

   The superior spatial resolution of {\sl HST} WFPC2 images combined with the
proximity of SBS 1415+437 permits to resolve individual stars and 
study stellar populations in this galaxy by means of
color-magnitude diagrams (CMD).
 We used the DAOPHOT routine PSF to fit the point spread
function and measure the brightness of individual stars. 
Figure 12 shows the distribution of photometric errors as a 
function of $V$ and $I$ magnitudes as determined by DAOPHOT.
We see that errors are of order 0.3 mag at 27 mag in $V$ and $I$, increasing 
to $\sim$ 0.5 mag at 28 mag. 


We have carried out a completeness analysis using the DAOPHOT routine 
ADDSTAR. For each frame we have added artificial stars amounting to $\sim$
10 \% of the real stars detected in each magnitude bin in the original image.
We then performed a new photometric reduction 
using the same procedure as the one applied to the original frame, and checked 
how many added stars were recovered. This operation was repeated 10 -- 20
times for each frame and the results were averaged. 
The completeness factor in each magnitude bin 
defined as the percentage of recovered artificial stars is shown 
in Table 5.
It can be seen that the completeness limit is reasonably good, 83 \% and 
70 \% respectively at $V$, $I$ $\sim$ 25 mag,
but drops to 22 \% and 4 \% at 27 mag.        
The total numbers of recovered stars in the frame 
are respectively 3677, 5652 and 1020 in 
the F569W band, the F791W band and in both of these bands at the same time.  

    We decided not to construct a CMD for the 
entire galaxy as is usually done   
(e.g. Dohm-Palmer et al. 1997ab, Schulte-Ladbeck, Crone \& Hopp 1998). 
Rather,  because the morphology of SBS 1415+437 and spectroscopic
observations suggest propagating star formation from
the NE to the SW, we have divided the BCD into 6 regions labeled
from I to VI as shown in Figure 13, and constructed separate CMDs 
for each of these regions. The total number of stars in the 6 CMDs is 785.  
 The F569W and F791W magnitudes are measured within a 2-pixel radius 
circular aperture, and then converted to magnitudes within a 0.5 arcsec 
radius circular aperture using the correction factors of Holtzman et al.
 (1995a).

The correction for internal extinction poses a problem. 
The BCD is undergoing an intense burst of star formation, and dust is 
expected to be produced in expanding envelopes of supernovae. We have already
noted that the ($V-I$) color distribution in the brightest part of the BCD
is inhomogeneous (Figure 7) implying the presence of dust patches.
Dust is known to be present even in the most metal-deficient BCDs.
A striking example is that of the BCD SBS 0335--052 with a metallicity 
of only 1/41 solar. It shows strong mid-infrared emission, implying a 
silicate dust mass of $\sim$ 10$^4$ $M_\odot$ (Thuan, Sauvage \& Madden 1999b). 
 Dust patches mixed in the star-forming region can be 
clearly seen on {\sl HST} pictures of SBS 0335--052 (Thuan et al. 1997), 
and spectroscopic observations
 (Izotov et al. 1997a) show the extinction to be high and to vary spatially.
Spatial variations of the extinction can also be seen in SBS 1415+437. 
Spectral observations of the brightest H II region (region III)
in different apertures show $E(B-V)$ varying from 0.01 to 0.08 mag (Table 2), 
on a scale less than 240 pc. 
The extinction is even larger ($E(B-V)$ = 0.18 mag or $A_V$ = 0.6 mag)
in other parts of the BCD (subsection 5.3). These spatial variations of 
the extinction introduce uncertainties in the derived CMDs. 
As a first approximation, we 
correct the CMD for each region by a constant internal extinction, when the  
spectral data are good enough to allow extinction estimates from the observed
Balmer decrements (see subsection 5.3). 
The CMD for region III  has been corrected for extinction corresponding to  
$A_V$ = 0.25 mag and $A_I$ = 0.15 mag, while the CMDs for regions IV -- VI 
have been corrected for extinction corresponding to 
$A_V$ = 0.55 mag and $A_I$ = 0.41 mag.
The CMDs for regions I and II have not been corrected for internal extinction 
for lack of information. 
In those CMDs, stars are fainter and redder than they really should be. 
Therefore, the comparison of the location of
stars with isochrones in those CMDs gives only upper limits on their ages. 
   
The $(V-I)$ vs. $I$ diagrams for each region are shown in Figure 14 together
with stellar isochrones by Bertelli et al. (1994)
for a heavy element abundance of 1/20 the solar value. Each isochrone is 
labeled by the logarithm of the age in yr. In Figures 14 c-f, 
the region occupied by 1 Gyr or older asymptotic giant branch (AGB) stars is
shown by dashed lines while the observational limits 
are shown by a dotted line. For the latter,
we adopt the magnitude limits $V$ = $I$ = 28 mag as suggested by
Figure 12. None of the stars lie beyond this line as they should.

   We now discuss the ages (or their upper limits) of the stellar populations 
  in each region. 
 The small regions I and II lie outside the main body of
the BCD to the SW of the brightest H II region, and contain
 two young stellar clusters.
Inspection of the $(V - I)$ color maps in Figures 7 and 8 shows that 
they have different ages. While red supergiants are present in region II, 
they are absent in region I which
is dominated by ionized gas indicative of the presence of hot O stars.
The CMDs confirm that scenario: 
 all stars in region I (Figure 14a) are still on the main sequence. 
 Region II is more evolved and has an age  
$\ga$ 10 Myr (Figure 14b). Region III contains the brightest H II region
(Figure 12). Its CMD (Figure 14c) shows a mixture of stellar populations 
with different ages, ranging from $\la$ 10 Myr to 100 Myr.
There appears to be red stars present in region III, 
with absolute magnitude $M_I$ in the
range --5.5 to --6.5 mag and ($V-I$) color in the range 1.0 to 2.0 mag.
However the colors of these red stars are quite uncertain: they have $I$
$\sim$ 25.5 mag, or $V$ $\sim$ 27.5 mag, so that their ($V-I$) colors can have 
errors as large as 0.4 mag. If real, 
these red stars can be either young AGB stars with age $\sim$ 100 Myr 
or red supergiant (RSG) stars (Gallart et al. 1996).     

Because of the distance of SBS 1415+437, the
imaging data presented here is not capable of detecting a 1 Gyr or older
 AGB stellar 
population, even if it exists. At $V$, $I$ = 26 mag, where the completeness is 
respectively 59 \% and 35 \% (Table 5), the detection limit for ($V-I$) = 2 mag is 
$I$ = 24 mag, nearly 1 mag above the AGB (upper dashed horizontal line 
in Fig. 14). By going down to $V$, $I$ = 27 mag which would 
allow to reach the AGB, the completeness limits drop precipitously to 
22 \% and 4 \%. Thus our CMDs cannot be used to exclude the presence of 
an old AGB stellar population. They can only provide a useful consistency 
check for stellar population ages derived by other techniques.    
Thus, the young ages inferred for the detected stars in region III from its CMD 
are in good 
agreement with those derived from the blue integrated $(V-I)$ color 
of --0.07 mag obtained by surface photometry (Table 5) and with those obtained  
from modeling the spectral energy distributions (subsection 5.3).

   The stellar populations in regions IV to VI are 
similar to those in region III except that the very youngest
population with age less 10 Myr is not present 
(Figures 14 d-f). 
Evolutionary synthesis models considered in 
subsection 5.3 give also an age $\la$ 100 Myr for regions IV to VI.
 From the CMD analysis, 
 there is clear evidence for 
 propagating star formation in SBS 1415+437 from the NE
(region VI) to the SW (region I). The CMDs show an evident trend of increasing 
age from region I to region VI. On smaller spatial scales, the same 
propagating star formation is seen within region III where the brightest
stars are distributed along a line in the  NE - SW direction
(Figure 6), with increasing age from the SW to NE (Figure 9).   

\subsection{Evolutionary synthesis models and the age of SBS 1415+437}

   Stronger evidence in favor of the evolutionary youth of SBS 1415+437
is provided by the spectral energy distributions (SED)
of regions III through VI as shown in Figure 15. 
All spectra are very blue.
The spectrum of region III (Figure 15a) contains many nebular 
emission lines: evidently, the continuum is dominated by the light of 
a very young O star population.
The spectra of the other regions (Figures 14 b-d) indicate older populations: 
their slopes are less steep and they show strong hydrogen
Balmer absorption lines characteristic of B and A stars. 

   To estimate quantitatively the age of each region, we calculate a grid 
of SEDs for stellar populations with ages varying between  
 10 Myr and 20 Gyr and with a heavy
element abundance $Z_\odot$/20, following the 
method described in Papaderos et al. (1998).
Briefly, we use isochrones of stellar parameters from
Bertelli et al. (1994) and the compilation of stellar atmosphere models from
Lejeune, Cuisinier \& Buser (1998). A Salpeter Initial Mass Function (IMF)
 with slope --2.35, an upper mass limit of 100 $M_\odot$ and a lower
mass limit of 0.6 $M_\odot$ were adopted. For stellar populations with
 age less than 10 Myr, we use the theoretical SEDs of  
Schaerer \& Vacca (1998) corresponding to a
 heavy element mass fraction $Z_\odot$/20
and a Salpeter IMF. The stellar emission in SBS 1415+437
is contaminated by ionized gas emission from supergiant H II regions.
Therefore, to study the stellar populations in SBS 1415+437 it is necessary
to construct a synthetic SED which includes both stellar and ionized gaseous
emission. We have chosen not to calculate the contribution of ionized gas
emission from the model value for the Lyman continuum luminosity. 
Instead, we add the
gaseous spectral energy distribution derived from the observed line
fluxes and equivalent widths, along with the observed emission lines 
in the spectral range
$\lambda$3700 -- 7500 \AA, to the calculated stellar spectral energy
 distribution. The gaseous emission contribution is
scaled to the stellar emission by the ratio of the observed H$\beta$ 
equivalent width to the one expected
for pure gaseous emission. The contribution of bound-free, free-free, 
two-photon continuum emission has been taken into account for the
spectral range from 0 to 5 $\mu$m (Aller 1984; Ferland 1980). 
The effect of gaseous emission is important in region III, but is minor in 
the other regions as indicated by their small H$\beta$ equivalent width. 
For example, $EW$(H$\beta$) is only 18 \AA\ in region V. We therefore neglect
gaseous emission in regions IV - VI.

In order to compare the observed SEDs with the model SEDs, we need to
correct the former for extinction. 
 In many BCDs,
the internal extinction derived from Balmer decrements significantly exceeds 
the Galactic extinction.
This is the case for SBS 1415+437 where both imaging and spectroscopic data
show the presence of significant internal extinction in the brightest H II 
region.

 In Figure 16a the observed spectrum of region III (thin line) is shown along
with the gaseous + synthetic stellar continuum produced by an 
instantaneous burst with age 4.7 Myr (thick line).
 The observed spectrum has been corrected for extinction, corresponding to 
 $C$(H$\beta$) = 0.12 dex or $E(B-V)$ = 0.08 mag as derived from the Balmer 
decrement. The agreement between the observed and model SEDs is very good,
except in the $\lambda$3600 - 3900 region where the model falls 
below the observed spectrum.
The discrepancy is
caused by the overlapping of the hydrogen Balmer absorption lines in the 
model SED. The inclusion of the observed Balmer hydrogen emission lines  
reduces the difference as shown in Figure 16b, where 
 the contributions of gaseous and stellar emission  are shown separately.
It is evident from Figure 16b that the contribution of ionized gas emission to
the total SED is significant in star-forming regions. This contribution is
dependent on the equivalent width of H$\beta$ and the observed fluxes of other
emission lines relative to H$\beta$. We present in Table 5 the  
model colors of a single stellar population with age 4.7 Myr. 
The calculated and observed $(V-I)$
colors agree very well. Thus the bulk of the stellar population in region III 
is very young ($\sim$ 5 Myr), although stars as old as 100 Myr are also present,
as shown by the CMD of this region (Fig. 14c). However, the
contribution of the older stars to the total light is negligible.

  Extinction correction is more of a problem  
when hydrogen emission lines are not present in the spectrum. Regions IV, V
and VI are characterized by nearly identical spectral energy distributions
(Fig. 15).
However, only in region V are hydrogen emission lines present with a 
good enough signal-to-noise ratio to allow for the
correction of internal extinction. Therefore only in that region 
can we derive the age of the stellar population. 

To illustrate the effect of extinction on the derived age, we consider 
in turn both the extinction-uncorrected and corrected cases. As the 
reddening for the stellar continuum can be different (and is usually 
smaller) than that derived from the Balmer lines in the ionized gas (e.g.
Calzetti, Kinney \& Storchi-Bergmann 1994), the real situation lies somewhere between these two limiting cases. However, the following arguments show that
the truth is closer to the case with extinction derived from the Balmer lines
than to the case with no extinction.
Figure 17a shows the spectrum of region V uncorrected 
for extinction, along with model SEDs
for three different ages log $t$ (yr) = 7.5, 8.0 and 8.1. 
The large difference between the model SEDs at log $t$ = 8.0 and 8.1 is due
to the appearance of the first AGB stars in the latter case. The best fit
 is obtained for the model with log $t$ = 8.1.
We present in Table 6 the model colors calculated for a single
stellar population with log $t$ varying between 7.2 and 8.1. 
It can be seen that the calculated $(V-I)$ color of 0.79 mag
 for the age log $t$ = 8.1 is 
inconsistent with the color of $\sim$ 0.4 derived 
from the HST images (this is the color of 
the flat part of the ($V-I$) color profile in Fig. 11b)
and uncorrected for interstellar
extinction. This is not too
surprising as the extinction derived for region V from the Balmer decrement
 is high: $C$(H$\beta$) = 0.26 dex, equivalent to 
$A_V$ = 0.55 mag and $A_I$ = 0.41 mag. 
Figure 17b shows the extinction-corrected spectrum of region V with  
four model SEDs corresponding to log $t$ = 7.2, 7.5, 8.0 and 8.1. The best
agreement between observation and model is achieved for a SED with log $t$
between 7.2 and 8.0. These ages give a range of   
model $(V-I)$ colors between 0.21 and 0.43, which does include the  
observed color of 0.25, corrected for the interstellar extinction. 
Interpolating between the two colors gives log $t$ = 7.25 (see also Table 6). 
 The above analysis shows the 
importance of extinction correction for age determination. Had we not 
corrected for extinction we would have derived an age log $t$ = 8.1 
instead of log $t$ = 7.3, i.e we would have overestimated the age by 
some 60 Myr. 
Extinction correction in region V  makes stars brighter 
by 0.6 mag in $V$ and by 0.4 mag in $I$, and red stars lie in
the young supergiant region rather in the AGB region (Fig. 14e).

We stress here that evolutionary synthesis is a very sensitive technique 
for determining ages of stellar populations. We have used it to obtain  
ages of 5 and $\sim$ 20 Myr for regions III and V, respectively. 
As the SEDs for regions IV and VI are very similar to that of region V
(Fig. 15), it is likely that their stellar populations have ages of 
$\sim$ 20 -- 100 Myr. Thus, the analysis of stellar populations in SBS 1415+437 
using evolutionary synthesis models has also led us to the conclusion
that the galaxy is young. Star formation in the BCD did not start until 
$\sim$ 100 Myr ago. This finding supports the conclusion of Izotov \& Thuan 
(1999) based on the analysis of chemical abundances in a large sample of 
BCDs that all galaxies with metallicities $\la$ $Z_\odot$/20 are young,
with ages less than $\sim$ 100 Myr.

\section {SUMMARY AND CONCLUSIONS}

We have analyzed the properties and examined the evolutionary status
of the nearby extremely low-metallicity ($Z_\odot$/21) blue compact dwarf (BCD) galaxy 
SBS 1415+437. Analysis of a very high signal-to-noise ratio long-slit 
MMT spectrum and of new {\sl HST} FOS spectroscopy and WFPC2 
imaging of the BCD has led us to the conclusion that
SBS 1415+437 is a nearby {\sl young} dwarf galaxy 
experiencing now its first burst of star formation which did not start until 
$\sim$ 100 Myr ago.
In particular, we have obtained the following results:

1. The oxygen abundance in SBS 1415+437 derived in two different apertures
is very similar: 12 + log (O/H) = 7.60 $\pm$ 0.01 and
7.61 $\pm$ 0.01, respectively, i.e. only 1/21 of that of the Sun
(adopting 12 + log (O/H)$_\odot$ = 8.93 from Anders \& Grevesse 1989).
This is in excellent agreement with the value  
12 + log (O/H) = 7.59 $\pm$ 0.01 derived earlier by 
Izotov \& Thuan (1998b) and in 
fair agreement with the value 12 + log (O/H) =
7.54 $\pm$ 0.14 derived from the lower signal-to-noise ratio {\sl HST}
FOS spectrum. 

2. The $\alpha$-elements-to-oxygen abundance ratios Ne/O, S/O, Ar/O
 in SBS 1415+437 are in very good agreement with the mean ratios
for the most metal-deficient blue compact galaxies derived by 
Izotov \& Thuan (1999).
For the largest aperture, we derive log (Ne/O) = --0.75 $\pm$ 0.02 as 
compared to the mean value --0.75 $\pm$ 0.03, log (S/O) = --1.58 $\pm$ 0.02 
as compared to the 
mean value --1.59 $\pm$ 0.04, and log (Ar/O) = --2.30 $\pm$ 0.03 as compared to 
the mean
value --2.22 $\pm$ 0.07. All these ratios are close to the respective solar
ratios implying a common origin of $\alpha$-elements in massive stars.

3. The nitrogen-to-oxygen abundance ratio 
log (N/O) = --1.54 $\pm$ 0.03 is in good agreement with the 
mean value of log (N/O) = --1.60 $\pm$ 0.02 derived by Izotov
\& Thuan (1999) for the most metal-deficient BCDs.
The very small spread of the N/O ratio in the most metal-deficient BCDs
($Z$ $\la$ $Z_\odot$/20) suggests that nitrogen 
in these galaxies is a {\sl primary} element made only by {\sl massive} stars.
Intermediate-mass stars in these very metal-deficient BCDs have not had yet  
time to evolve and release their nucleosynthesis products. This implies that 
these galaxies, including SBS 1415+437, are not older than $\sim$ 100 Myr
(Izotov \& Thuan 1999).  

4. The iron-to-oxygen abundance ratio [O/Fe] = 0.27 $\pm$ 0.03 is in excellent 
agreement with the mean value of 0.32 $\pm$ 0.11 
derived by Izotov \& Thuan (1999) for the most metal-deficient BCDs, and 
is similar to the values found in Galactic halo stars. 
Iron is produced only during explosive nucleosynthesis either by
short-lived massive stars, progenitors of supernovae of type II or by
long-lived low-mass stars, progenitors of supernovae of type I. 
Oxygen is overproduced relative to iron in SBS 1415+437 as compared to the Sun,
suggesting that iron is produced by short-lived massive stars. Since 
 SNe I start to explode after $\sim$ 1 Gyr after the beginning of star
formation, the [O/Fe] value allows us to put an upper limit of 1 Gyr
on the age of SBS 1415+437.

5. Carbon and silicon abundances have been derived using FOS
{\sl HST} observations. We find log (C/O) = --0.78 $\pm$ 0.10, in excellent
 agreement with the mean value of --0.78 $\pm$ 0.03 found by 
 Izotov \& Thuan (1999) for the most metal-deficient BCDs and that of  
 --0.83 expected in the case of
carbon production by massive stars only (Weaver \& Woosley 1993).
As in the case of nitrogen, the C/O ratio implies that SBS 1415+437 is young 
enough so that intermediate-mass stars have not had time to evolve and 
release their C production. Again, this gives an upper limit for the age of 
SBS 1415+437 of $\sim$ 100 Myr.
  As for Si, we find log (Si/O) = --1.46 $\pm$ 0.27 close to the 
solar value of --1.38, as expected in the case of Si production in massive
stars. We do not find significant Si depletion onto dust grains, 
contrary to the 
conclusion of Garnett et al. (1995b). 

6. The heavy element abundances are constant within the largest supergiant
H II region (region III) with a diameter of $\sim$ 500 pc,
similar to the spatial distribution of heavy elements in other extremely
 metal-deficient galaxies such as I Zw 18 (Martin 1996) and SBS 0335--052
(Izotov et al. 1997a). To within the errors, there are also no 
heavy element abundance variations in different H II regions within
SBS 1415+437. This constancy strongly argues for H II region self-enrichment 
(Kunth \& Sargent 1986).
 
7. SBS 1415+437 is one of the best targets for the determination of the 
primordial helium mass fraction $Y_p$ because of its
very low metallicity, high surface brightness and
 low electron number density which
minimizes the influence of collisional enhancement of He I emission line
intensities. The helium mass fraction in SBS 1415+437 is $Y$ = 0.246 $\pm$ 0.004,
in very good agreement with the primordial
value $Y_p$ = 0.244 $\pm$ 0.002 derived by Izotov \& Thuan (1998b).

8. {\sl HST} WFPC2 $V$ and $I$ images of SBS 1415+437 show a comet-like 
structure with a very blue bright supergiant H II region at its SW
tip, and extended diffuse emission along the elongated main body of the galaxy 
on which are superposed red supergiant stars.
The $V$ and $I$ surface brightness profiles are well fitted by exponentials
with a very small scale-length $\alpha^{-1}_V$ = $\alpha^{-1}_I$ = 0.19 kpc 
and a central surface brightness $\mu_{V_0}$ = 21.00 mag arcsec$^{-2}$ and 
$\mu_{I_0}$ = 20.69 mag arcsec$^{-2}$. The $(V-I)$ color of the brightest H II
region is very blue $\la$ 0. The galaxy becomes 
gradually redder with increasing 
distance from the brightest H II region, reaching a constant 
$(V-I)$ $\sim$ 0.4 at $r$ $\ga$ 5\arcsec. These color changes imply changes in
stellar population ages and propagating star formation in the BCD from the NE
 to the SW. Comparison of the observed colors with 
 those of evolutionary synthesis models gives an upper limit for the 
 age of the stellar populations of $\sim$ 100 Myr.    
The resolved bright stars in the brightest
H II region are also aligned along the NE - SW direction, supporting the 
hypothesis of the sequential character of star formation in SBS 1415+437. 
The $(V-I)$ color image shows a filamentary structure 
caused by the superposition of many supernova supershells mixed in with
 dust patches. The H$\alpha$ and [O III] $\lambda$5007 velocity
distributions imply a solid-body rotation of the galaxy along the minor axis 
with a rotational velocity of 80 km s$^{-1}$. The dynamical mass
within the optical diameter is 1.9 $\times$ 10$^9$ $M_\odot$ or $\sim$ 13 times
larger than neutral gas mass, indicating that the major part of the 
galaxy's  mass is in the form of dark matter.

9.  $(V-I)$ - $I$ color-magnitude diagrams (CMD) have been constructed 
for the stellar populations in 6 separate regions along the body of 
SBS 1415+437. We use MMT spectral observations to derive and  
correct for extinction for regions III to VI (Fig. 12), the spectra of 
regions I and II not showing strong enough Balmer lines to derive extinction.
We have adopted as a first approximation a constant extinction in each region, 
although there is evidence that the extinction varies spatially on scales less 
than $\sim$ 240 pc.  
The CMDs show a gradual increase of stellar population ages from the SW to NE,
supporting the hypothesis of propagating star formation. 
SBS 1415+437 is too far away
($D$ = 11.4 Mpc) for 1 Gyr or older asymptotic giant
branch stars to be detected. 
 Red stars are seen however with $(V-I)$ = 1 -- 2 mag, although these colors are 
subject to large errors ($\sim$ 0.4 mag). 
Corrected for extinction, their mean absolute
magnitude $M_I$ is $\sim$ --5.5 to --6.5 mag, which makes them, if real, to be   
young red supergiant or young bright asymptotic giant branch stars 
with age $\la$ 100 Myr. 

10. MMT long-slit spectroscopy along the major axis of SBS 1415+437
shows that everywhere in the galaxy, the spectral energy distribution (SED) is
blue. We calculate model SEDs including both stellar and ionized gas emission 
for different regions in the galaxy, and found that the best fit for 
the brightest H II region gives a stellar age of $\sim$ 5 Myr while  
the ages of the stellar populations in the other regions of the galaxy do not 
 exceed 100 Myr.

    In conclusion, chemical abundances, color profiles and 
evolutionary synthesis models for the spectral energy distribution 
all give an age of $\la$ 100 Myr for SBS 1415+437. 
The BCD is truly a young galaxy undergoing now its 
first burst of star formation. Along with I Zw 18 ($Z_\odot$/50, 
Hunter \& Thronson 1995),
SBS 0335--052 ($Z_\odot$/41, Thuan et al. 1997, Izotov et al. 1997a), 
SBS 1415+437 ($Z_\odot$/21) is the third extremely 
metal-deficient BCD, analyzed in detail for the age of its stellar populations,
which confirms the conclusion of Izotov \& Thuan (1999) that all galaxies 
with metallicities $\la$ $Z_\odot$/20 are young, with age not exceeding 
$\sim$ 100 Myr.

\acknowledgements
Y.I.I. thanks the staff of the Astronomy Department of the University of 
Virginia for their kind hospitality. We thank Polis Papaderos for his help in 
deriving the surface brightness profiles.
Partial financial support for this 
international collaboration was made possible by NATO collaborative research 
grant 921285 and NSF grant AST-9616863.


\clearpage

\figcaption[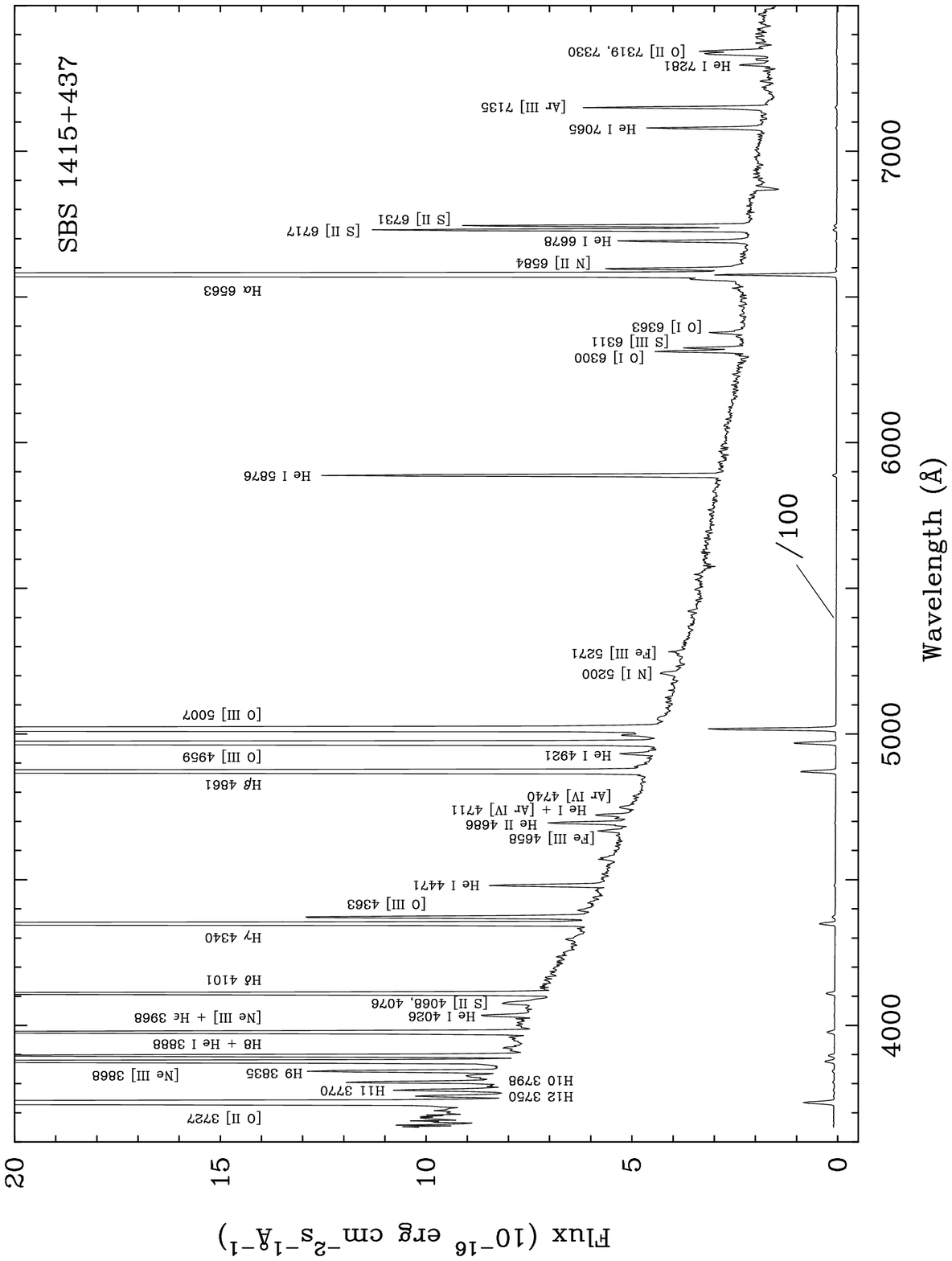]{MMT spectrum of SBS 1415+437 within a 
1\farcs5 $\times$ 5\arcsec\ aperture with nebular emission line identifications.}

\figcaption[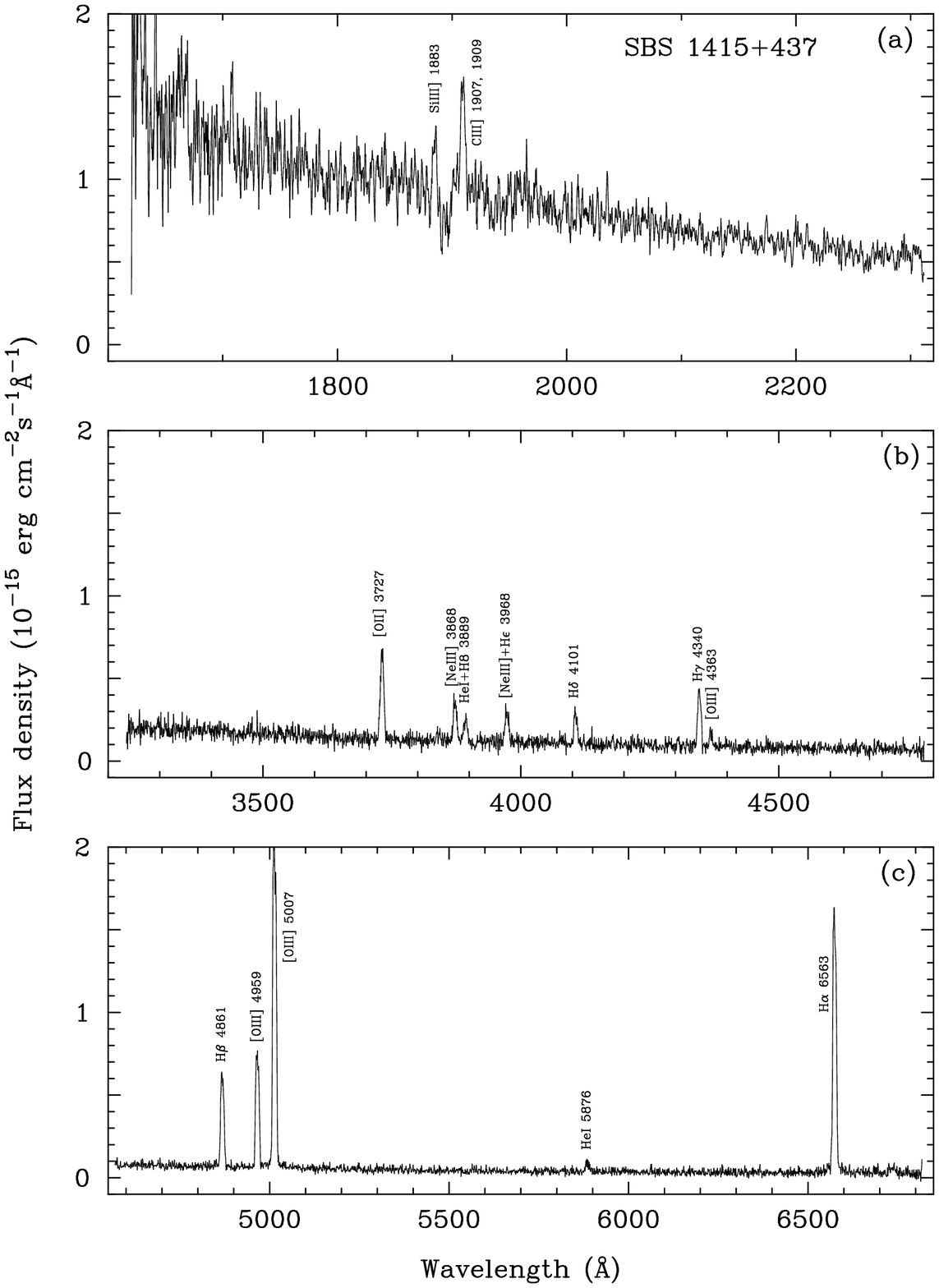]{{\sl HST} FOS spectrum of SBS 1415+437 
within a 0\farcs86 round aperture with nebular emission line 
identifications. The Si III] $\lambda$1892 emission line is not present in
the UV spectrum. Instead an absoption feature is seen.}

\figcaption[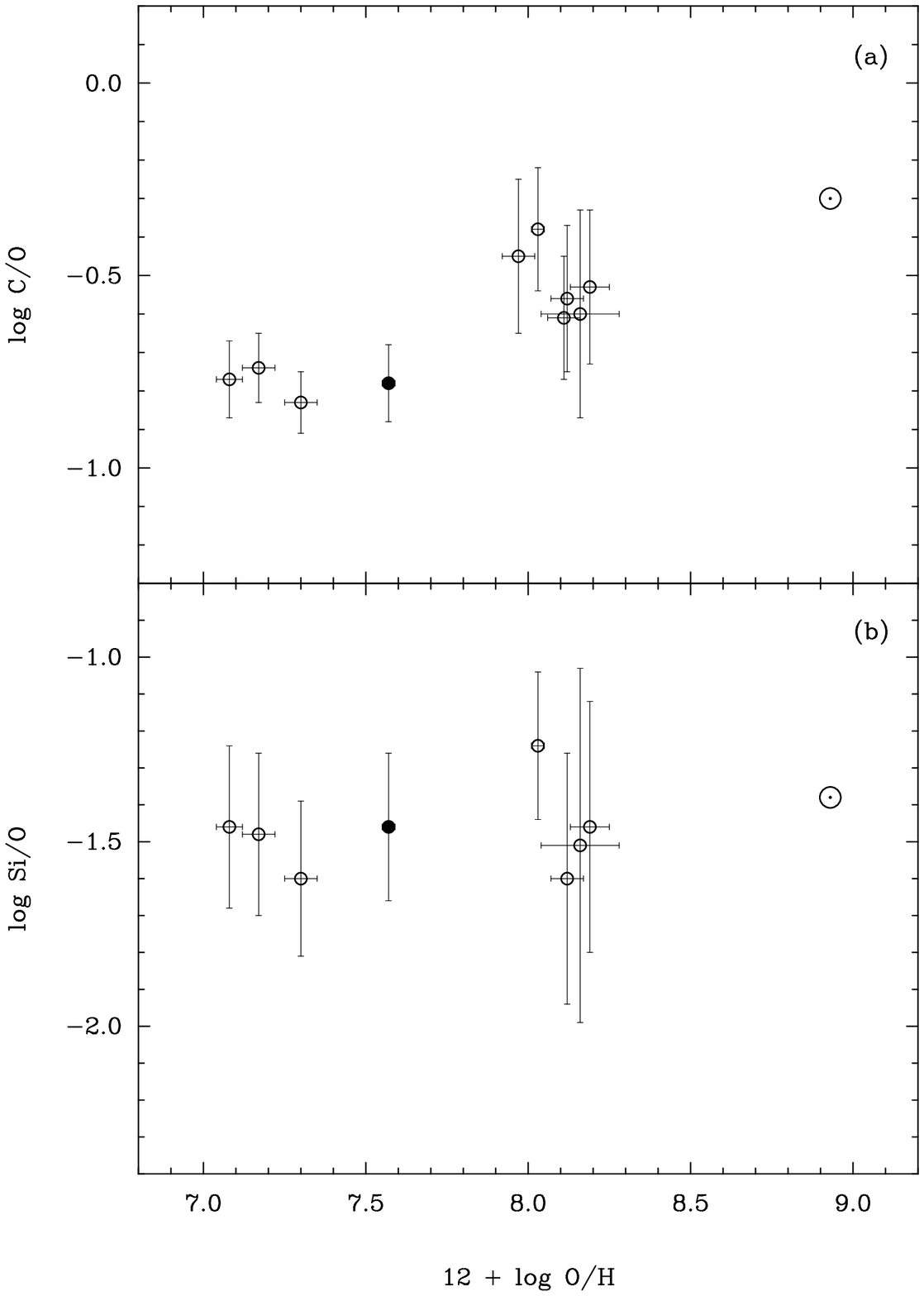]{ (a) The C/O abundance ratio vs. oxygen abundance 
for a sample of metal-deficient blue compact galaxies (open circles, 
Izotov \& Thuan 1999).
The filled circle represents SBS 1415+437. The solar value is shown by 
the symbol $\odot$;
(b) Si/O abundance ratio vs. oxygen abundance. The symbols have the same 
meaning as in Fig. 3a. }

\figcaption[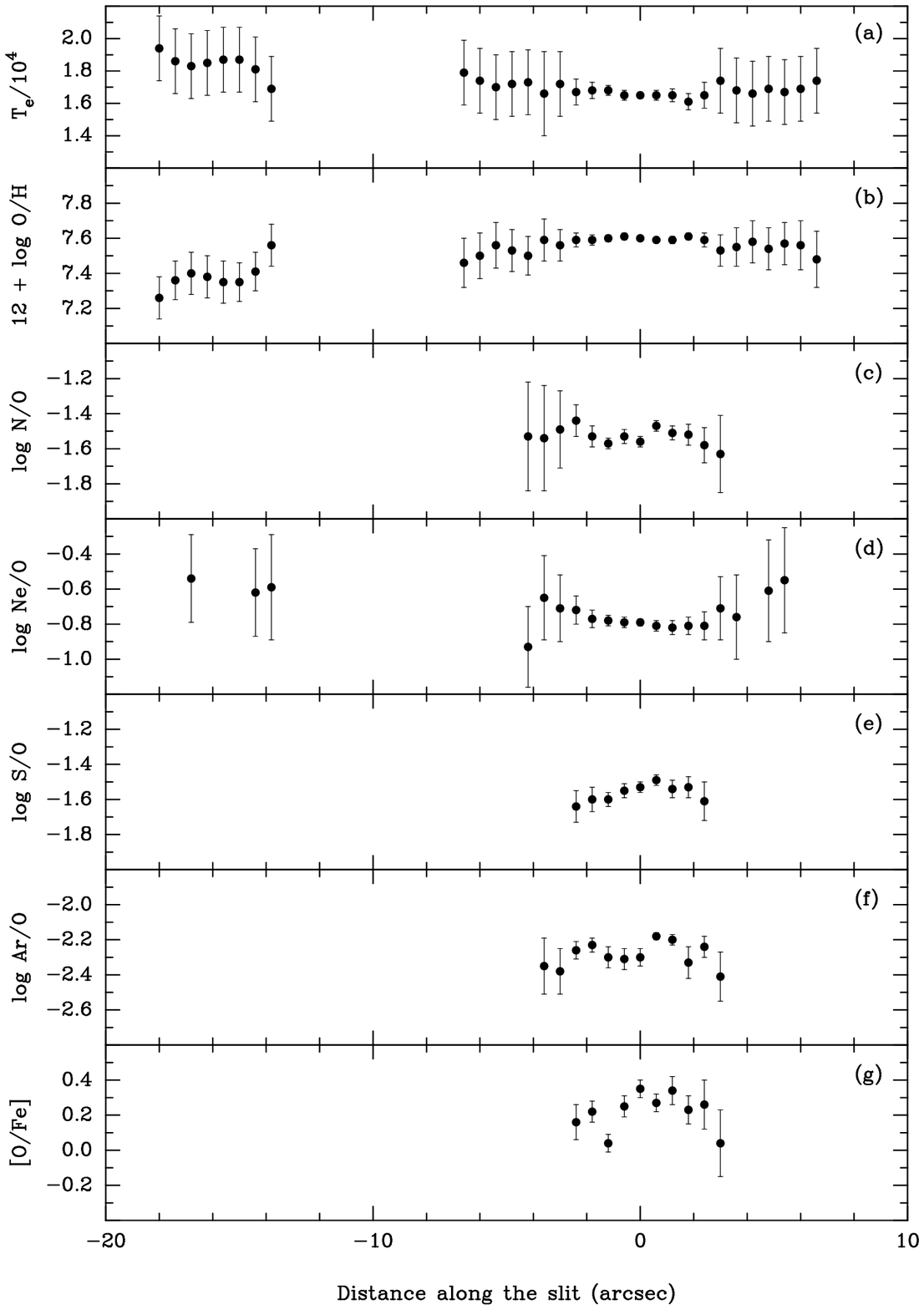] {Spatial distribution of the electron temperature $T_e$(O 
III), oxygen abundance and heavy element-to-oxygen abundance ratios. The origin 
is taken to be at the brightest H II region. At the distance $D$ = 11.4 Mpc of
SBS 1415+437, 1\arcsec\ is equivalent to 55 pc.}

\figcaption[fig5.ps]{{\sl HST} WFPC2 $I$ image of the comet-like 
galaxy SBS 1415+437 with 
a bright supergiant H II region at its SW side and unresolved diffuse stellar 
emission and resolved stars along the body 
of the galaxy.}

\figcaption[fig6.ps]{Magnified $I$ image of the brightest H II region 
in SBS 1415+437. There is a chain of bright stars along the NE - SW direction
implying propagating star formation.}

\figcaption[fig7.ps]{{\sl HST} $(V-I)$ image of SBS 1415+437. 
Blue $(V-I)$ colors are dark and red $(V-I)$ colors are light.
There are three blue H II regions at the SW end of the galaxy and 
two fainter H II regions in its central part.    
The brightest H II region has a filamentary structure 
and shows supernova superbubbles with dust patches mixed in.
Resolved red stars can be seen in the main body of the galaxy. }

\figcaption[fig8.ps]{Magnified $(V-I)$ image of the brightest H II region.
Blue $(V-I)$ colors are dark and red $(V-I)$ colors are light.  
The white points to the NE are resolved red stars.
There is a clear age difference for the two stellar clusters to the SW.
 The southernmost cluster contains only hot O stars and is younger while
red supergiant stars are present in the older cluster to the north.}

\figcaption[fig9.ps.ps]{$(V-I)$ color vs. distance for
stars in the brightest H II region. There is a clear color gradient, implying 
a systematic age gradient from the NE to the SW and propagating star 
formation.} 

\figcaption[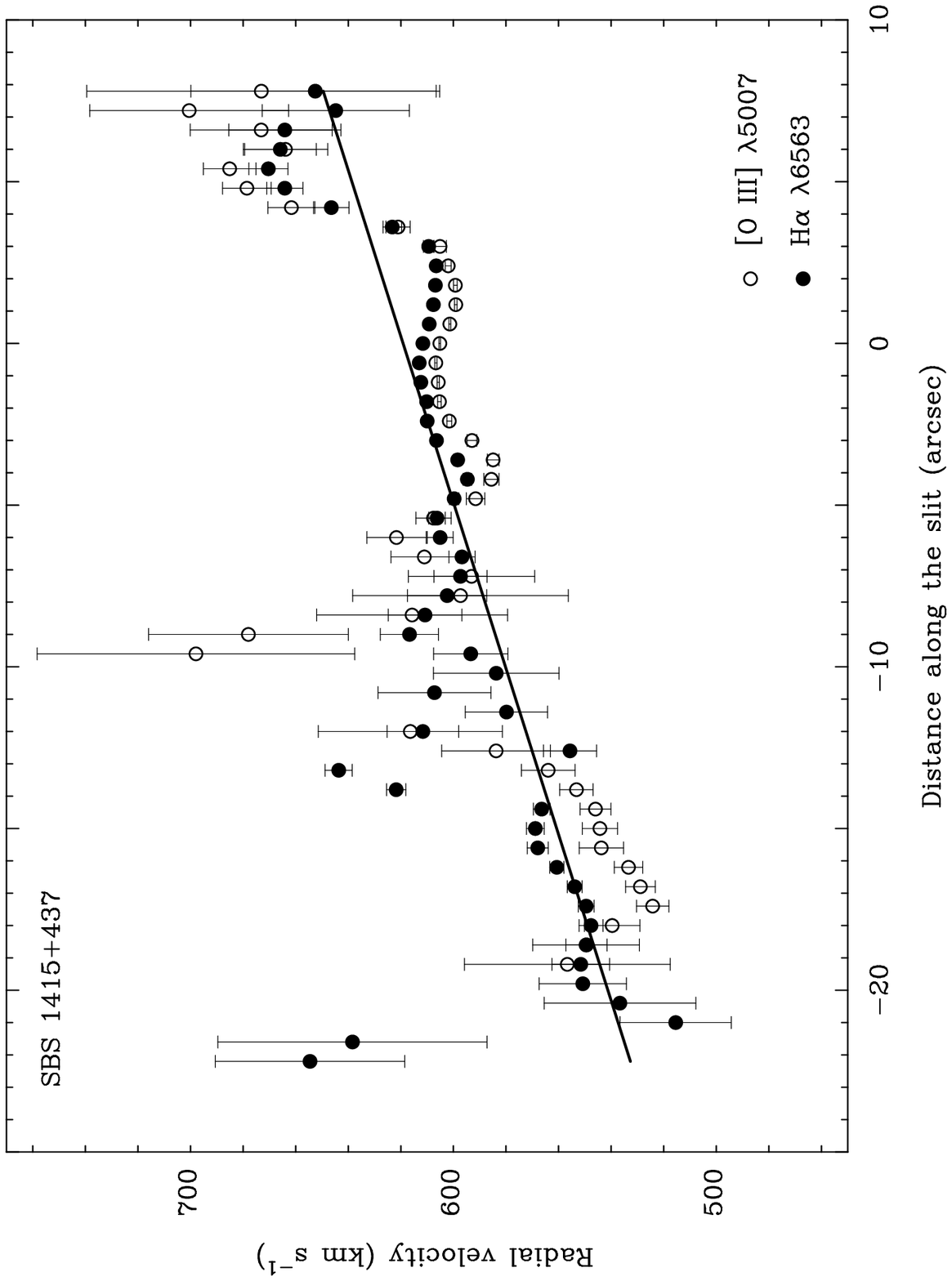]{Spatial distribution of the ionized gas velocity as 
derived from the H$\alpha$ 
(filled circles) and [O III] $\lambda$5007 (open circles) emission lines.
The origin is placed on the brightest H II region. The linear 
regression fit to the data points is shown by the solid line.} 

\figcaption[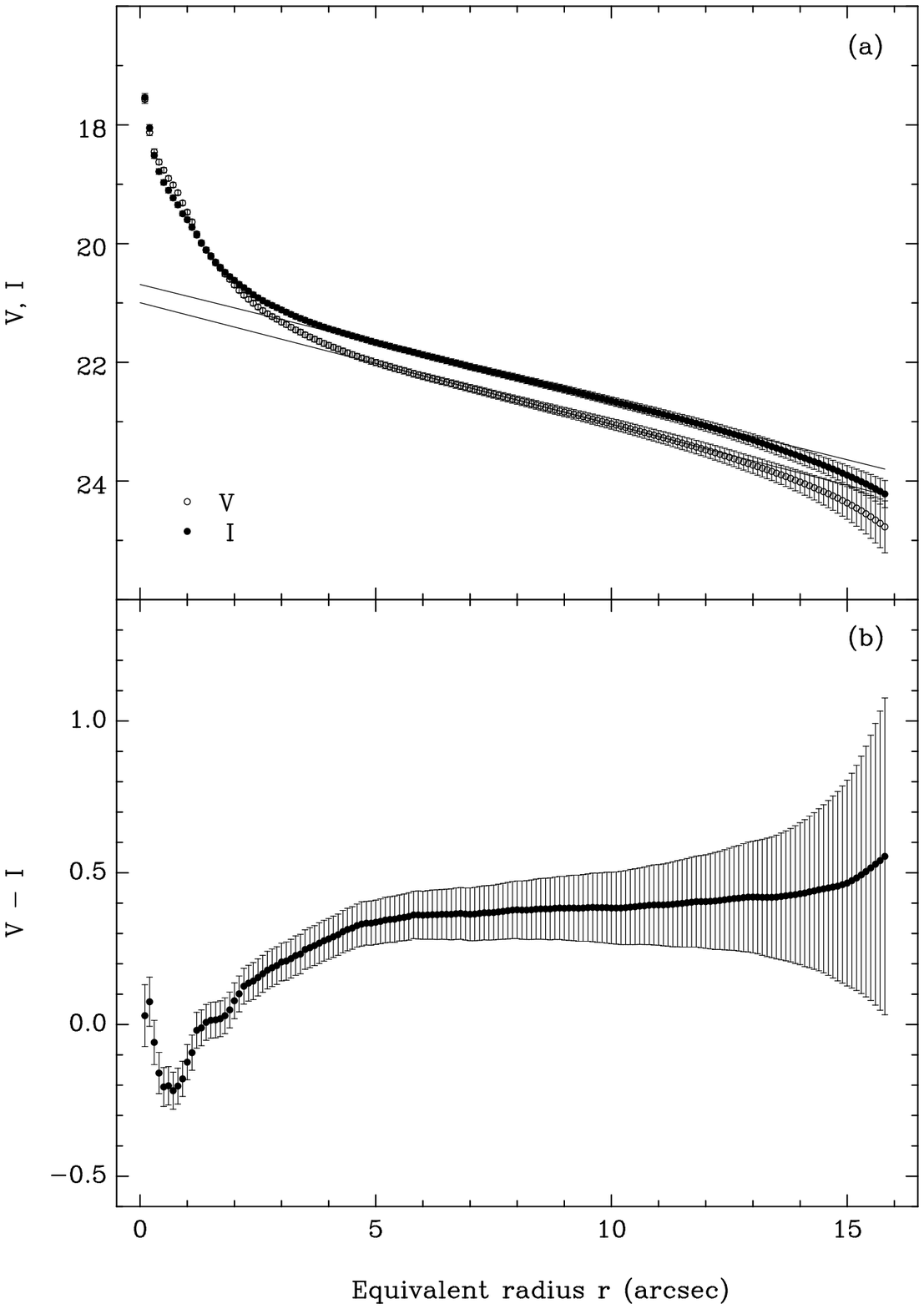]{(a) $V$ (open circles) and $I$ (filled circles) surface
brightness distributions as a function of the equivalent radius. The exponential
fits are shown by solid lines;
(b) $(V-I)$ color distribution as a function of the equivalent radius. The blue
colors at $r$ $\la$ 5\arcsec\ are associated with the star-forming region. The
diffuse low-intensity extended component has a nearly constant color.}

\figcaption[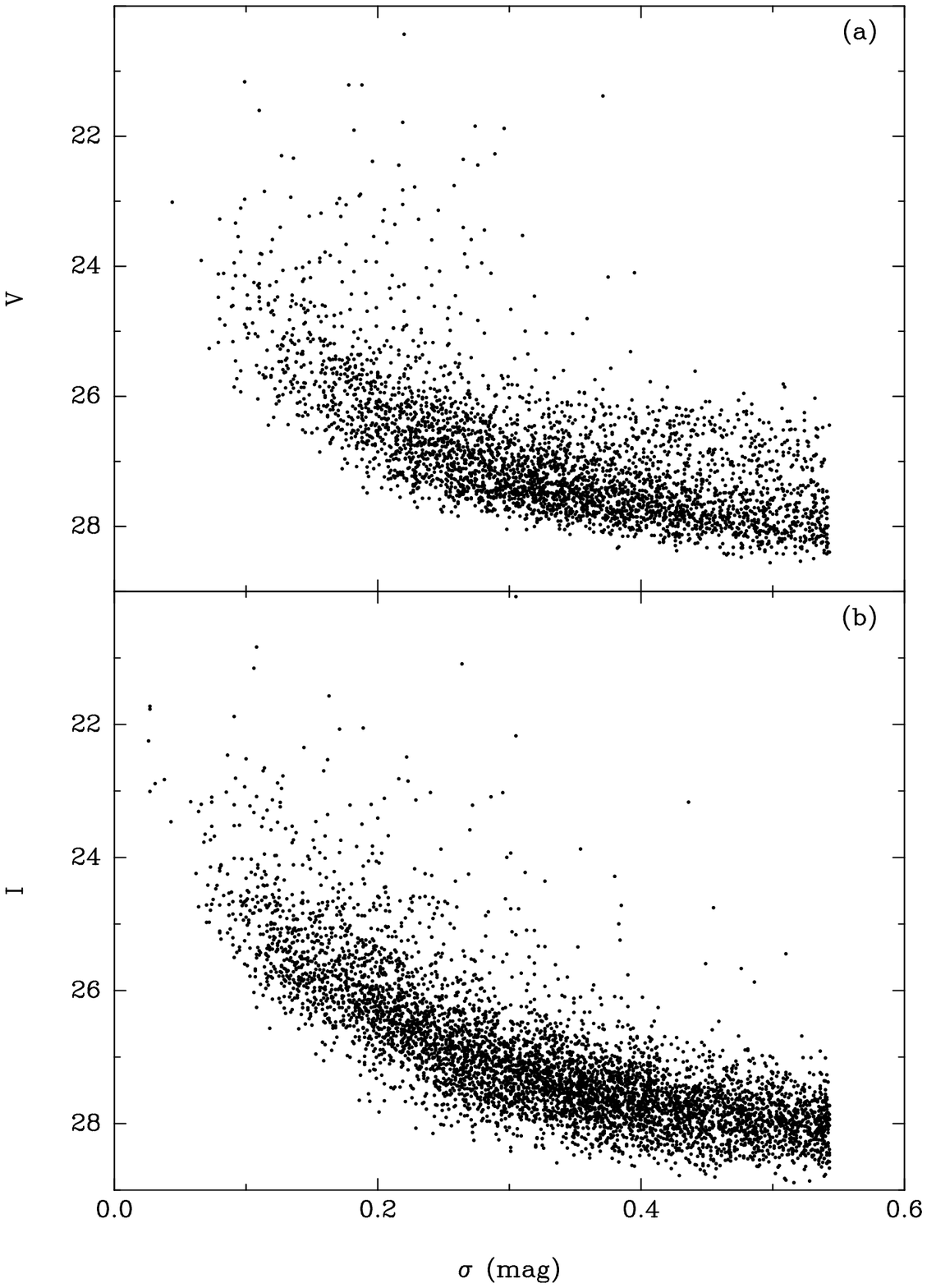]{The distribution of errors with magnitude as determined
by DAOPHOT for (a) $V$ and (b) $I$ photometry.}

\figcaption[fig13.ps]{Labeling of the 6 different regions in SBS 1415+437
 selected for stellar population color - magnitude 
diagram analysis.}

\figcaption[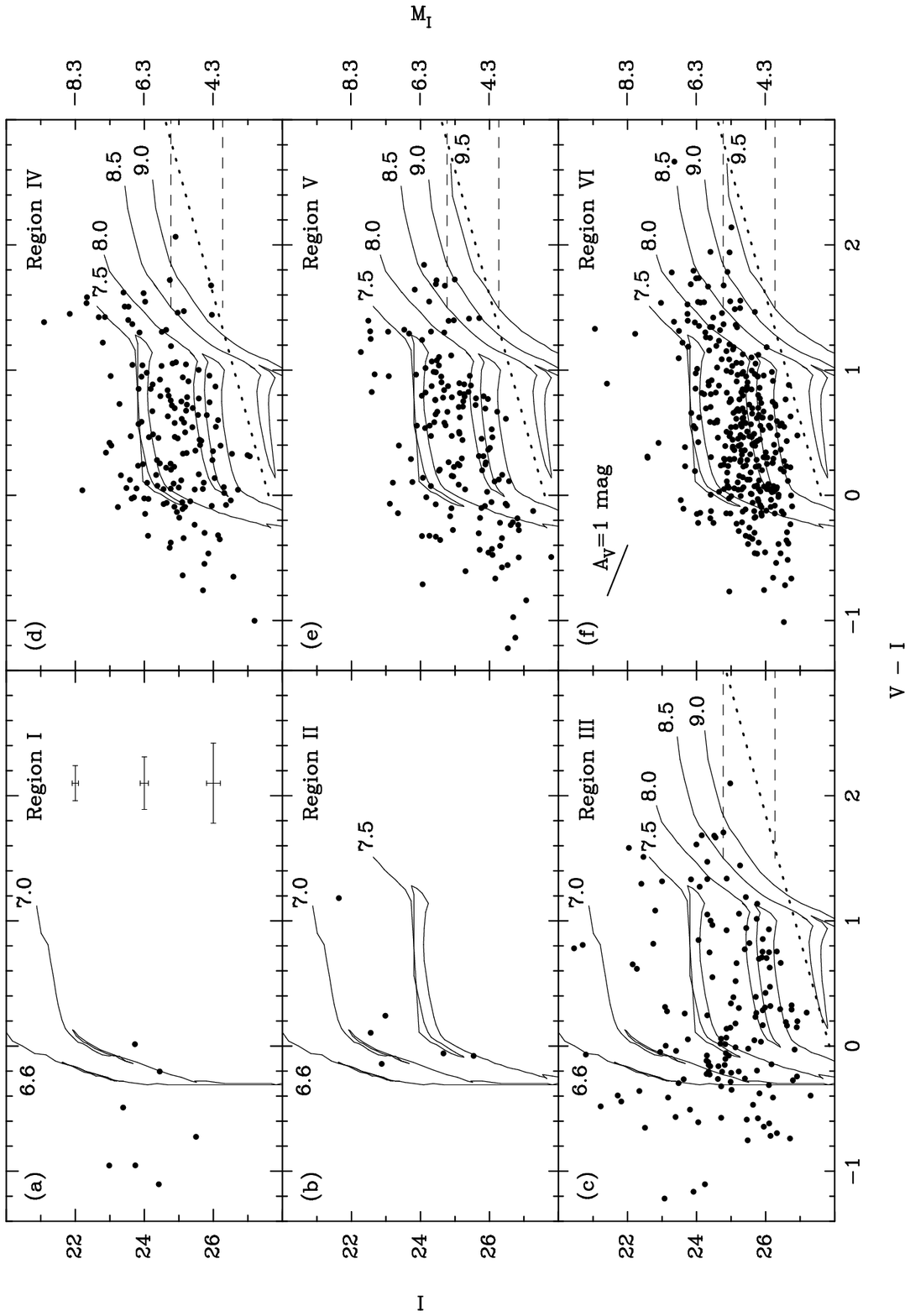]{Color - magnitude diagrams of stellar populations
in the regions of SBS 1415+437 defined in Figure 13. The 
isochrones (solid lines) are from Bertelli et al. (1994) and are labelled
by the logarithm of age in yr. Dashed lines represent the region
of 1 Gyr asymptotic giant branch stars. Dotted lines show the 
extinction-corrected observational limits corresponding to $V$ and $I$ equal to 
28 mag. There is a gradual 
age increase from region I to region VI, implying propagating star 
formation from the NE (region VI) to the SW (region I). The typical photometric
errors are shown in Fig. 14a.}

\figcaption[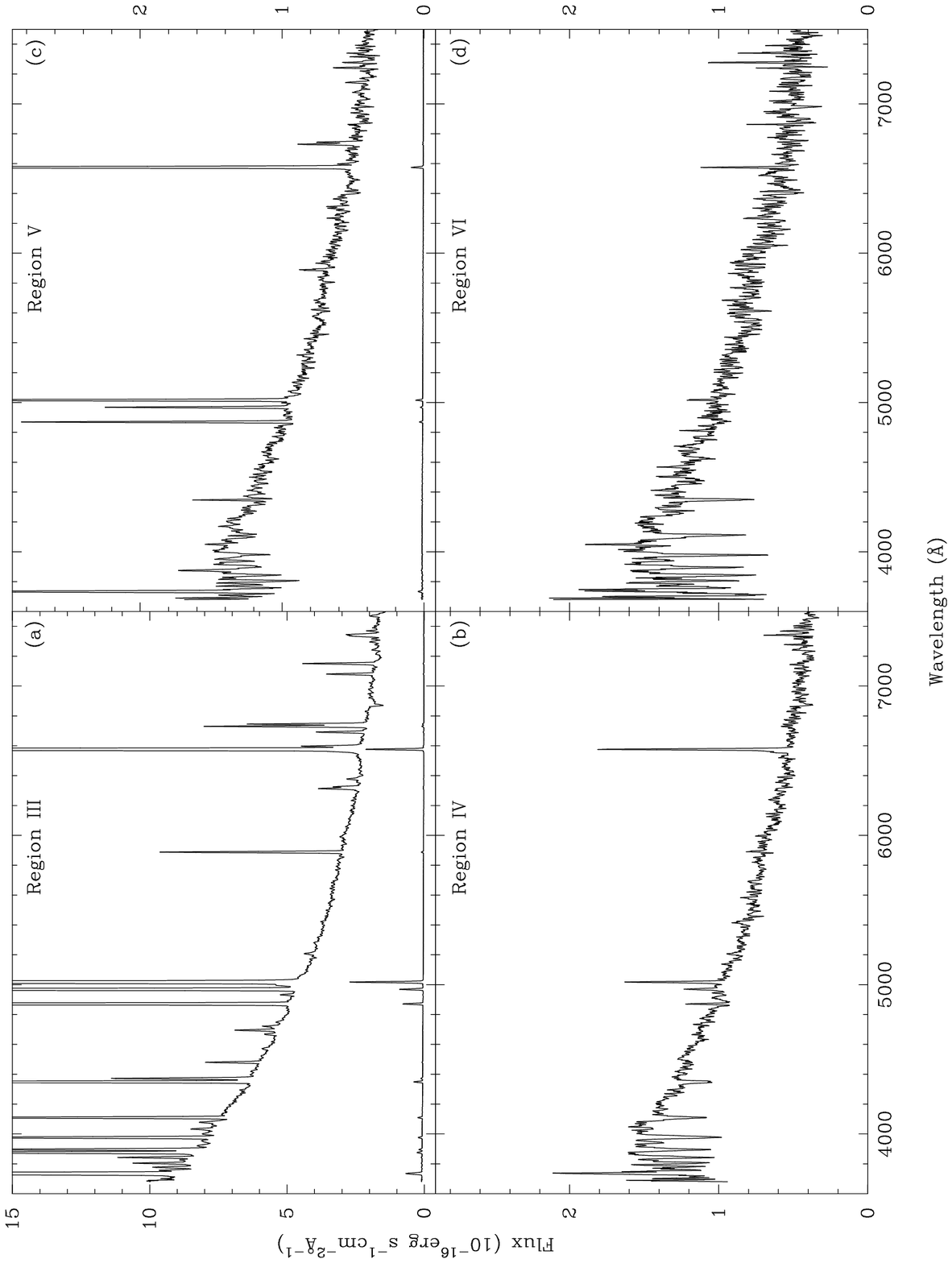]{MMT spectra of regions III to VI. The spectral 
energy distributions show blue colors indicative of young stellar populations.}

\figcaption[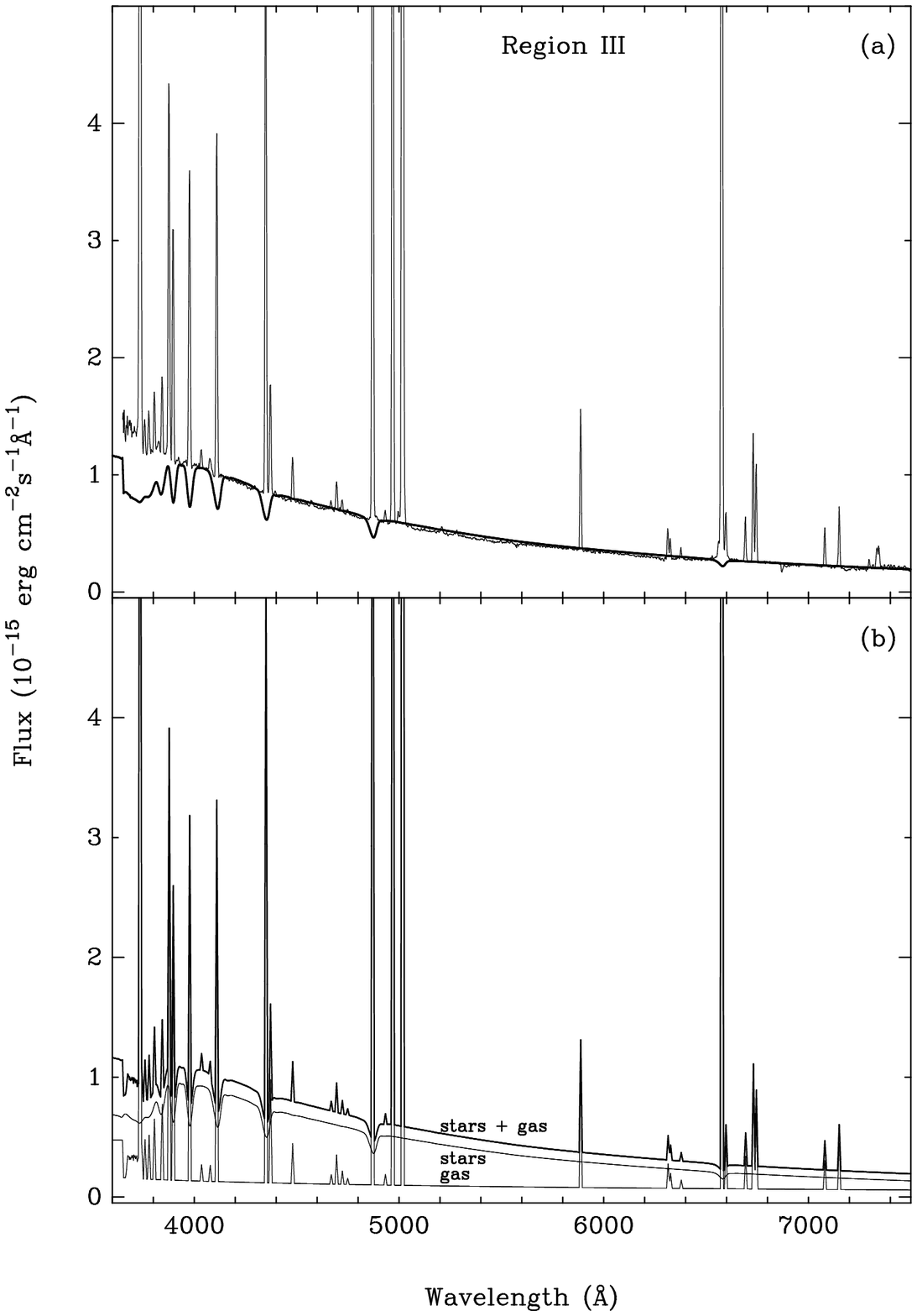]{(a) Extinction-corrected spectrum of region III 
(thin solid line) on which is superposed the theoretical 
stellar + gaseous spectral energy distribution (SED) 
of a stellar population with age 
4.7 Myr (thick solid line). 
The stellar energy distribution is taken from Schaerer \& Vacca
(1998) and the gaseous continuum is calculated using the 
observed equivalent width of H$\beta$ and electron temperature $T_e$(O III).
(b) Decomposition of the  
theoretical stellar + gaseous spectral energy distribution in Fig. 16a 
(thick solid
line) into stellar and gaseous emission. The relative
contribution of the gaseous emission increases toward the red part of 
the spectrum. The emission line fluxes are taken from the observations.}

\figcaption[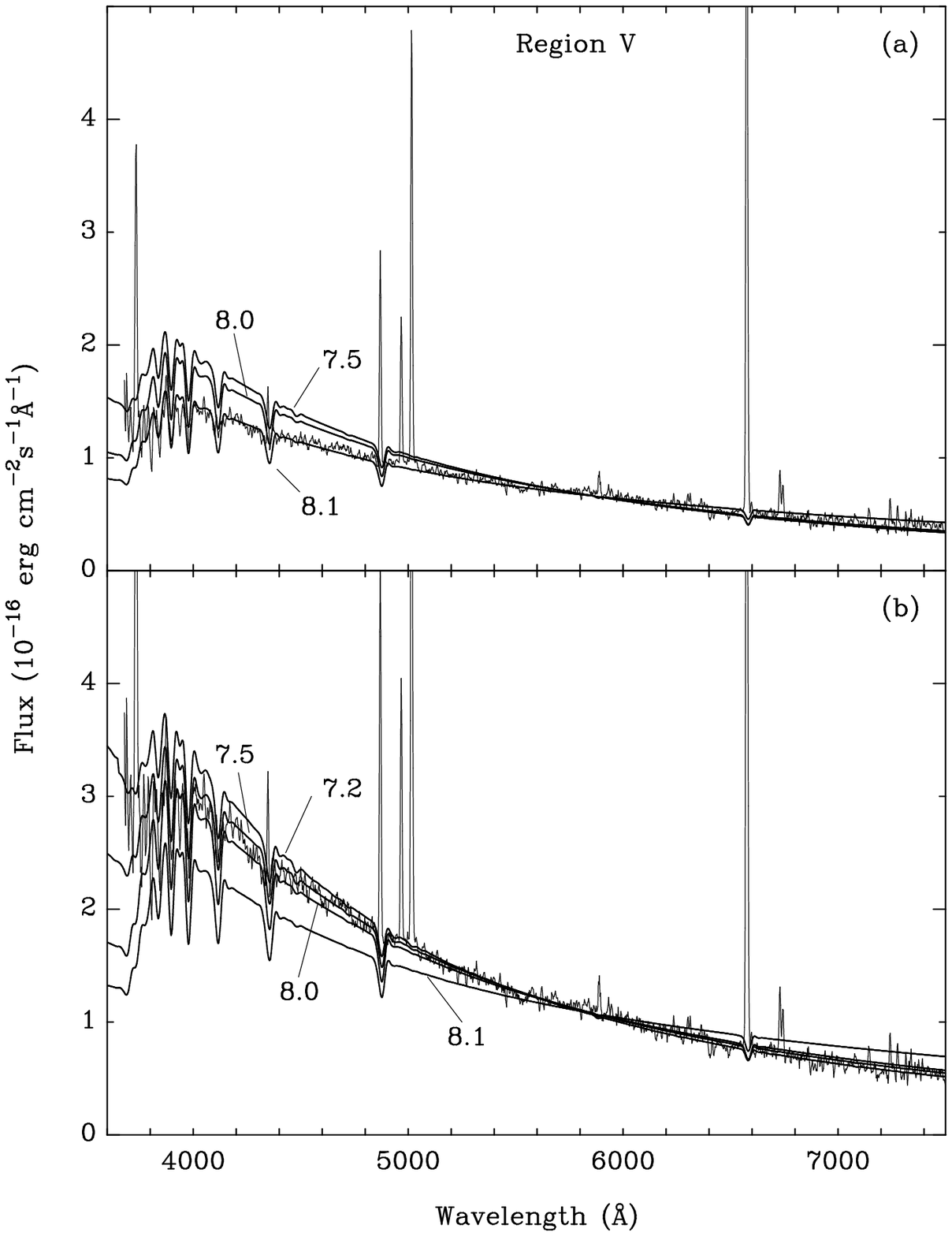]{ (a) Uncorrected for extinction spectrum of region
V on which are superposed the theoretical energy distributions of
 stellar populations
with ages log $t$ = 7.5, 8.0 and 8.1. The strong change in spectral energy
distributions (SED) between log $t$ = 8.0 and log $t$ = 8.1 is caused by the 
appearance of the first asymptotic giant branch stars in the latter case. 
The theoretical SEDs are calculated using
isochrones from Bertelli et al. (1994) and the stellar atmosphere model 
compilation
from Lejeune et al. (1998), for a metallicity equal to 1/20 that of the Sun. 
A Salpeter IMF and a lower stellar mass limit of 0.6 $M_\odot$ have been
adopted. The best fit is achieved with a theoretical SED with age log $t$ = 8.1;
(b) Extinction-corrected spectrum of region V on which are superposed 
theoretical SEDs.  The best fit is a SED with log $t$ between 7.2
and 8.0.}

\clearpage


\begin{figure*}
\figurenum{t1}
\epsscale{2.0}
\plotfiddle{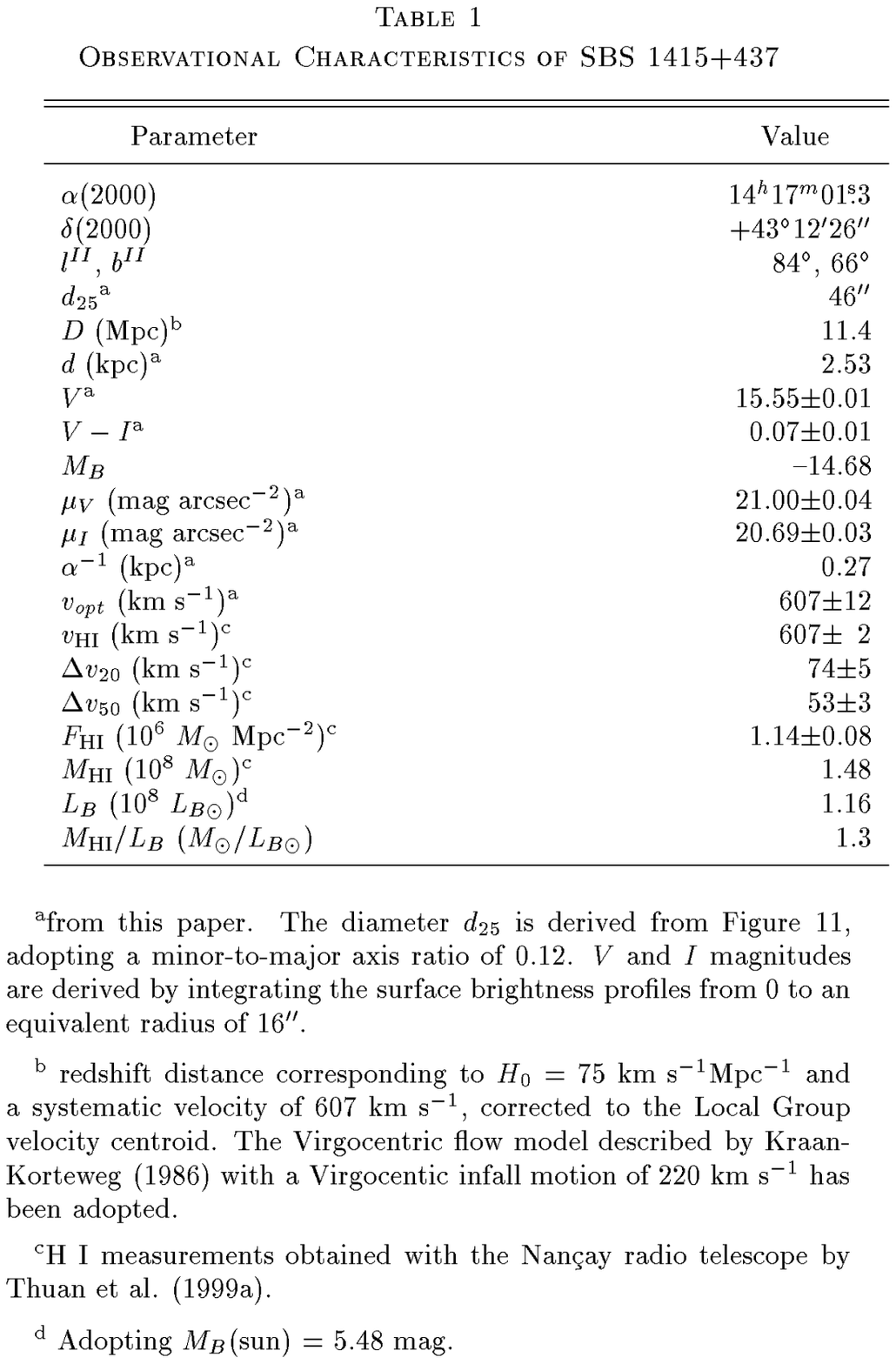}{0.cm}{0.}{100.}{100.}{-290.}{-350.}
\end{figure*}

\clearpage


\begin{figure*}
\figurenum{t2}
\epsscale{2.0}
\plotfiddle{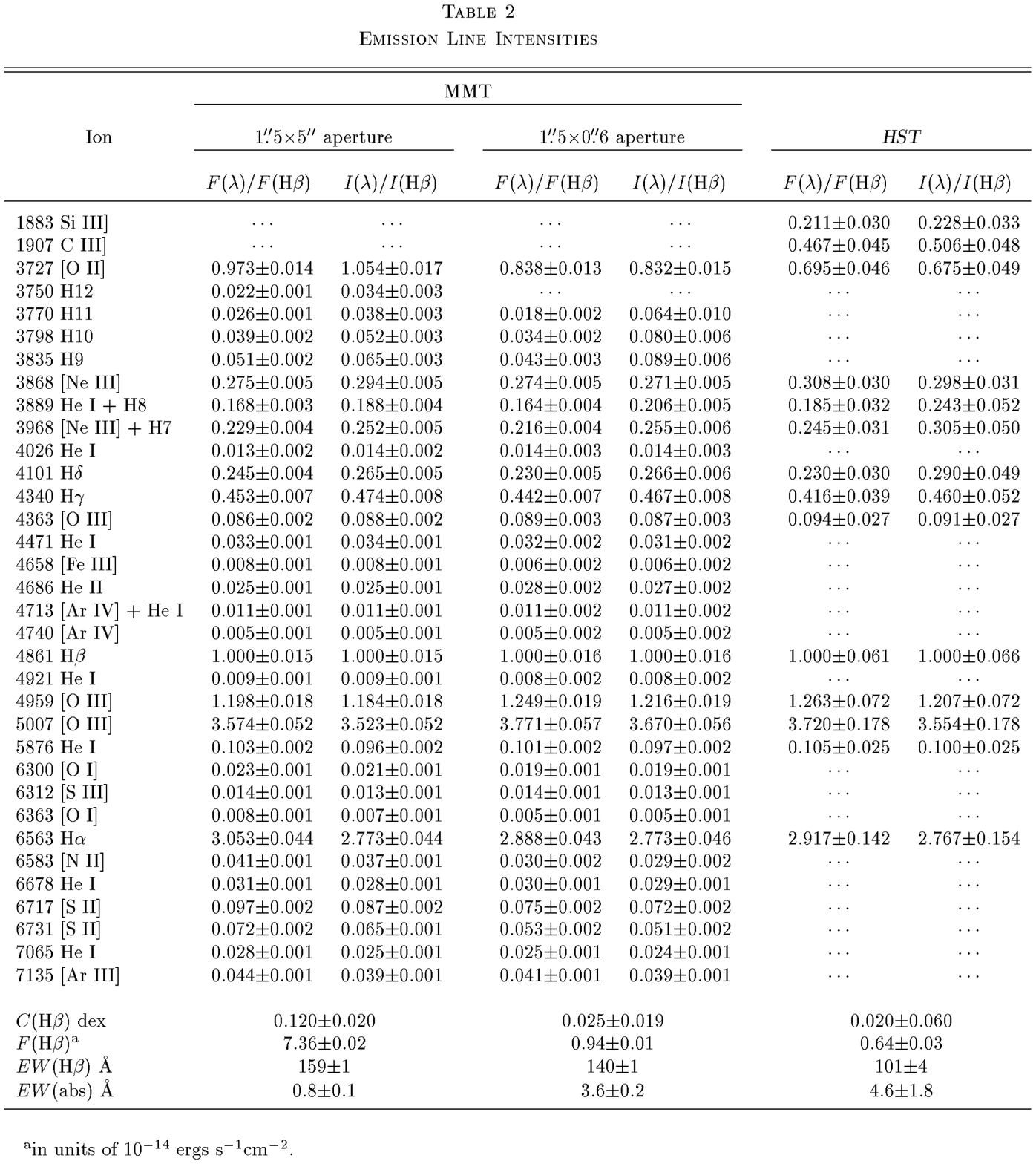}{0.cm}{0.}{95.}{95.}{-320.}{-350.}
\end{figure*}

\clearpage


\begin{figure*}
\figurenum{t3}
\epsscale{2.0}
\plotfiddle{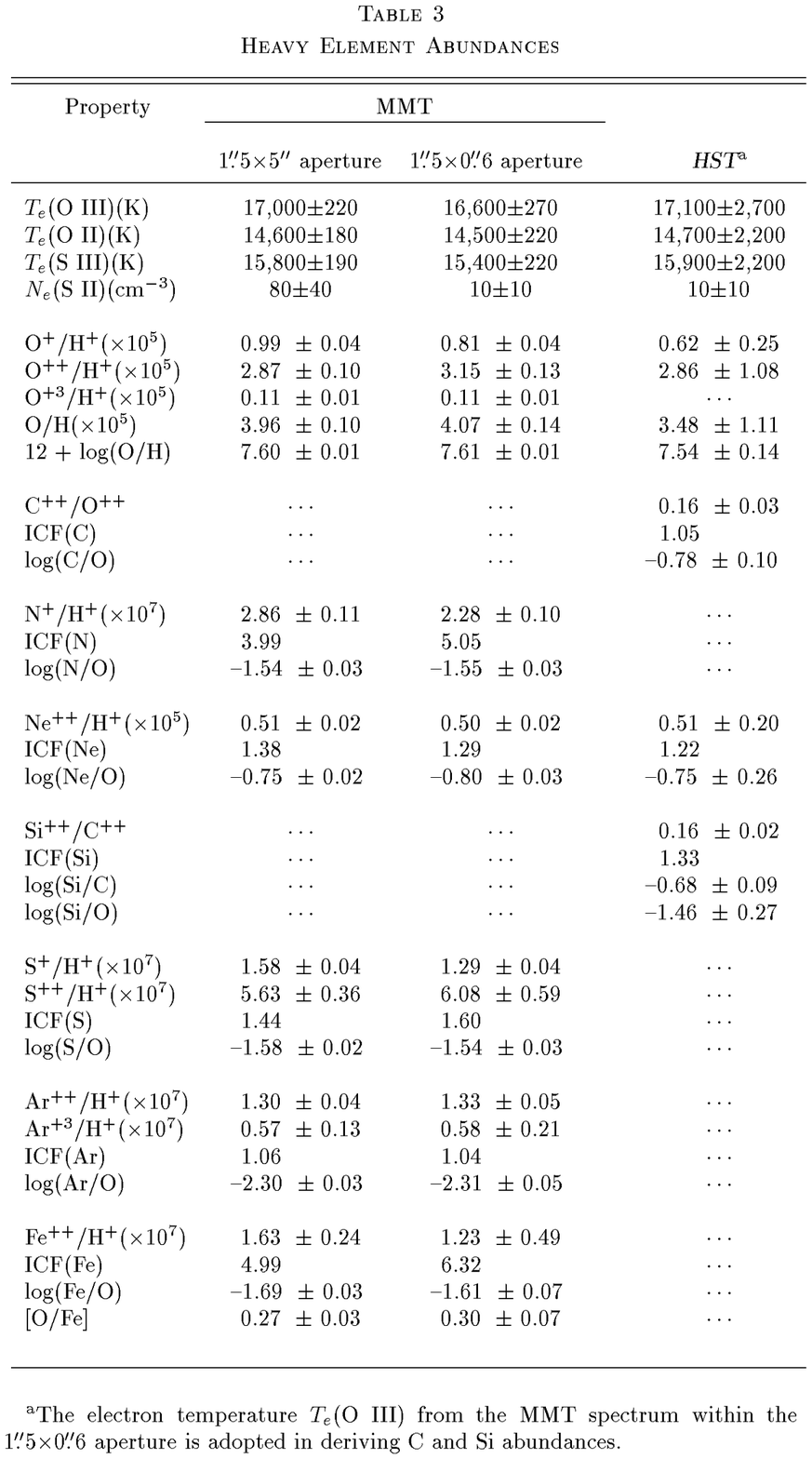}{0.cm}{0.}{100.}{100.}{-290.}{-350.}
\end{figure*}

\clearpage


\begin{figure*}
\figurenum{t4}
\epsscale{2.0}
\plotfiddle{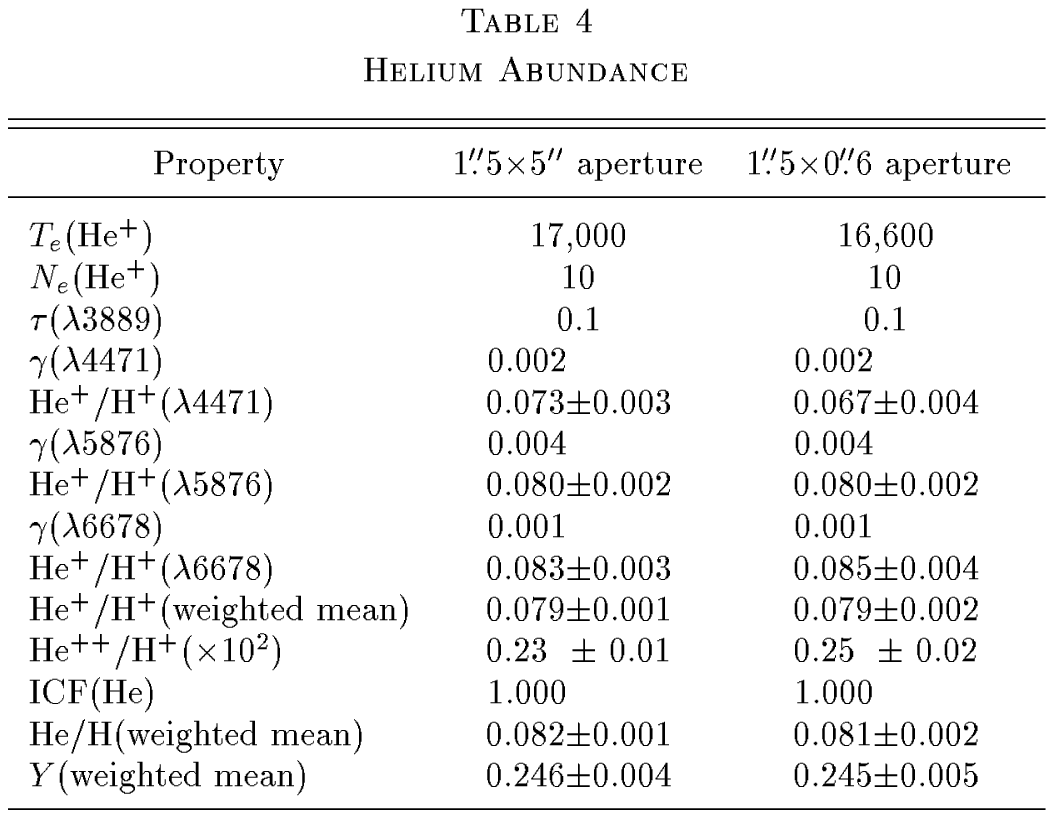}{0.cm}{0.}{100.}{100.}{-290.}{-350.}
\end{figure*}



\begin{figure*}
\figurenum{t5}
\epsscale{2.0}
\plotfiddle{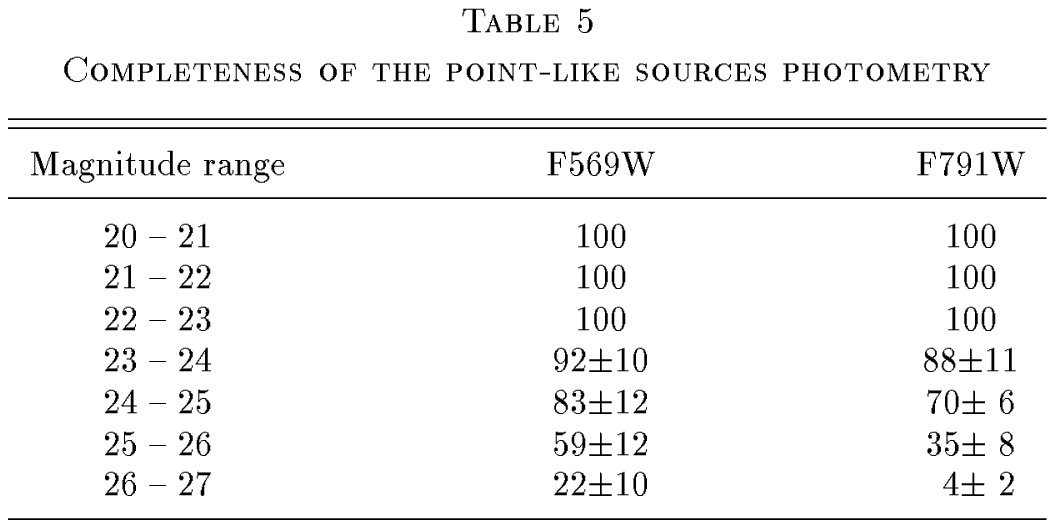}{0.cm}{0.}{100.}{100.}{-290.}{-350.}
\end{figure*}

\clearpage


\begin{figure*}
\figurenum{t6}
\epsscale{2.0}
\plotfiddle{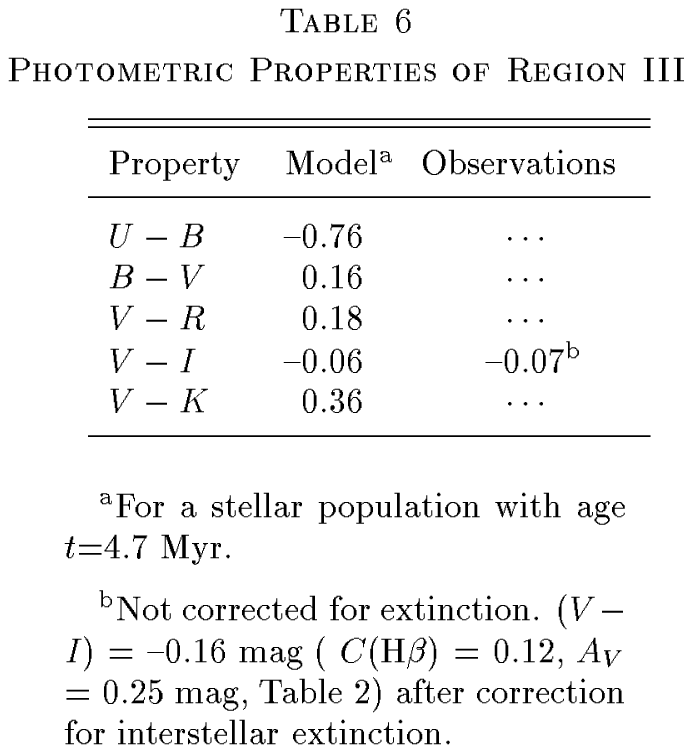}{0.cm}{0.}{100.}{100.}{-290.}{-350.}
\end{figure*}



\begin{figure*}
\figurenum{t7}
\epsscale{2.0}
\plotfiddle{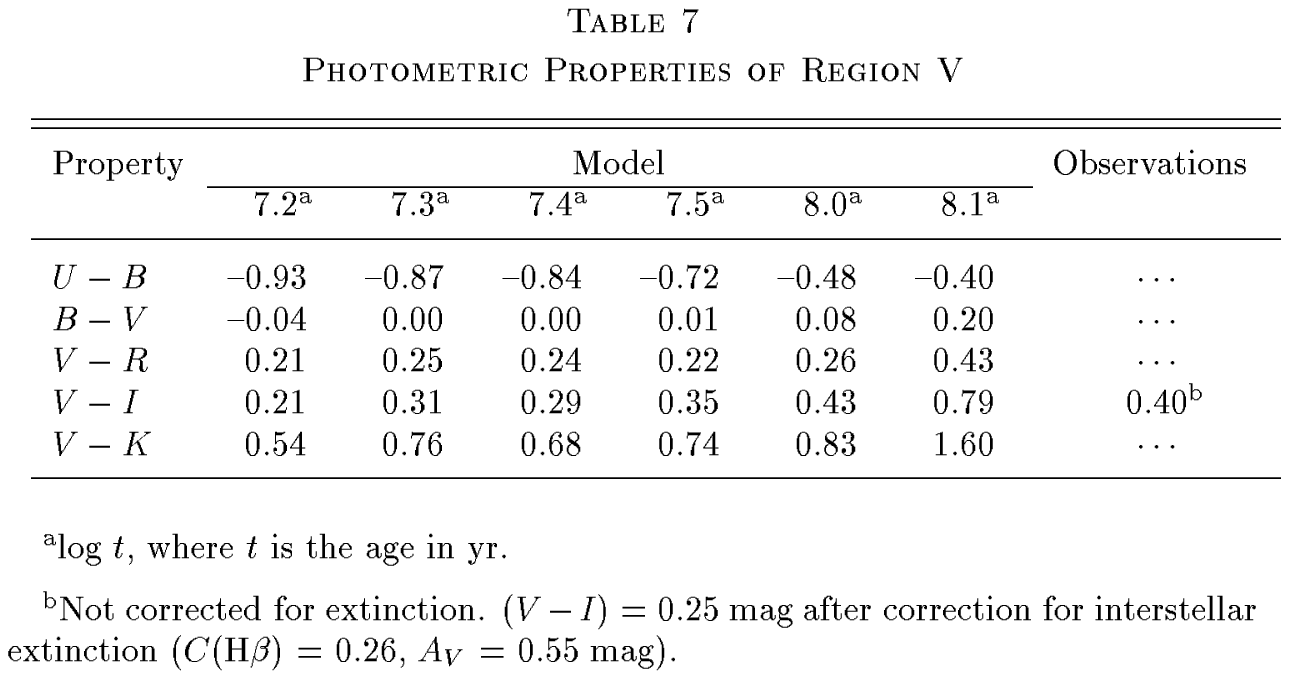}{0.cm}{0.}{100.}{100.}{-290.}{-350.}
\end{figure*}

\end{document}